\documentclass[11pt,a4paper]{article}

\usepackage{iftex}
\ifPDFTeX
  \usepackage[T1]{fontenc}
  \usepackage[utf8]{inputenc}
  \usepackage{lmodern}
\else
  \usepackage{fontspec}
\fi
\usepackage{amsmath,amssymb,amsthm}
\usepackage{mathtools}
\usepackage[margin=2.6cm]{geometry}
\usepackage{microtype}
\usepackage{enumitem}
\usepackage{booktabs}
\usepackage{xcolor}
\usepackage{tikz}
\usetikzlibrary{arrows.meta}
\usepackage[hidelinks]{hyperref}
\usepackage{lineno}

\newif\ifshowcomments \showcommentsfalse

\numberwithin{equation}{section}
\newtheorem{theorem}{Theorem}[section]
\newtheorem{lemma}[theorem]{Lemma}
\newtheorem{proposition}[theorem]{Proposition}
\newtheorem{corollary}[theorem]{Corollary}
\theoremstyle{definition}
\newtheorem{definition}[theorem]{Definition}
\newtheorem{construction}[theorem]{Construction}
\newtheorem{example}[theorem]{Example}
\theoremstyle{remark}
\newtheorem{remark}[theorem]{Remark}

\newcommand{\N}{\mathbb{N}}
\newcommand{\Z}{\mathbb{Z}}
\newcommand{\Par}{\mathcal{P}}
\newcommand{\reach}[1]{[#1\rangle}
\newcommand{\fire}[1]{\xrightarrow{\,#1\,}}
\DeclareMathOperator{\lcm}{lcm}
\newcommand{\NP}{\textsf{NP}}
\newcommand{\coNP}{\textsf{coNP}}
\newcommand{\Ptime}{\textsf{P}}
\newcommand{\pb}{\textup{\textsc{RingLive}}}
\newcommand{\cover}{\textup{\textsc{Cover}}}
\newcommand{\zero}{\mathbf{0}}

\pgfkeys{/ringpic/.is family,/ringpic/.cd,
  R/.initial=3.3, RL/.initial=4.35, RW/.initial=3.78, psize/.initial=11mm, theight/.initial=7.5mm,
  u0/.initial={$\bar r_0$}, v0/.initial={$W{-}\bar r_0$}, u1/.initial={$\bar r_1$}, v1/.initial={$W{-}\bar r_1$},
  u2/.initial={$\bar r_2$}, v2/.initial={$W{-}\bar r_2$},
  b0/.initial={$b_0$}, b1/.initial={$b_1$}, b2/.initial={$b_2$}, W/.initial={$W$}}
\newcommand{\ringRIW}[1]{%
  \begingroup
  \pgfkeys{/ringpic/.cd,#1}%
  \def\rR{\pgfkeysvalueof{/ringpic/R}}\def\rL{\pgfkeysvalueof{/ringpic/RL}}\def\rW{\pgfkeysvalueof{/ringpic/RW}}%
  \def\rk##1{\pgfkeysvalueof{/ringpic/##1}}%
  \begin{tikzpicture}[
    place/.style={circle,draw,thick,minimum size=\rk{psize},inner sep=0pt,font=\footnotesize},
    trans/.style={rectangle,draw,thick,fill=black!15,minimum width=2.4mm,minimum height=\rk{theight},inner sep=0pt},
    arc/.style={-{Latex[length=2.2mm]},thick},
    w/.style={font=\footnotesize,inner sep=1pt},
    lab/.style={font=\small}]
    \node[place] (u0) at (90:\rR)   {\rk{u0}};
    \node[trans,rotate=-30]  (a0) at (60:\rR) {};
    \node[place] (v0) at (30:\rR)   {\rk{v0}};
    \node[trans,rotate=-90]  (b0) at (0:\rR) {};
    \node[place] (u1) at (-30:\rR)  {\rk{u1}};
    \node[trans,rotate=-150] (a1) at (-60:\rR) {};
    \node[place] (v1) at (-90:\rR)  {\rk{v1}};
    \node[trans,rotate=-210] (b1) at (-120:\rR) {};
    \node[place] (u2) at (-150:\rR) {\rk{u2}};
    \node[trans,rotate=-270] (a2) at (-180:\rR) {};
    \node[place] (v2) at (150:\rR)  {\rk{v2}};
    \node[trans,rotate=-330] (b2) at (120:\rR) {};
    \draw[arc] (u0) -- (a0); \draw[arc] (a0) -- (v0);
    \draw[arc] (v0) -- (b0); \draw[arc] (b0) -- (u1);
    \draw[arc] (u1) -- (a1); \draw[arc] (a1) -- (v1);
    \draw[arc] (v1) -- (b1); \draw[arc] (b1) -- (u2);
    \draw[arc] (u2) -- (a2); \draw[arc] (a2) -- (v2);
    \draw[arc] (v2) -- (b2); \draw[arc] (b2) -- (u0);
    \node[w] at (75:\rW) {\rk{b0}};   \node[w] at (45:\rW) {\rk{b0}};
    \node[w] at (15:\rW) {\rk{W}};    \node[w] at (-15:\rW) {\rk{W}};
    \node[w] at (-45:\rW) {\rk{b1}};  \node[w] at (-75:\rW) {\rk{b1}};
    \node[w] at (-105:\rW) {\rk{W}};  \node[w] at (-135:\rW) {\rk{W}};
    \node[w] at (-165:\rW) {\rk{b2}}; \node[w] at (165:\rW) {\rk{b2}};
    \node[w] at (135:\rW) {\rk{W}};   \node[w] at (105:\rW) {\rk{W}};
    \node[lab] at (90:\rL) {$u_0$};    \node[lab] at (60:\rL) {$\alpha_0$};
    \node[lab] at (30:\rL) {$v_0$};    \node[lab] at (0:\rL) {$\beta_0$};
    \node[lab] at (-30:\rL) {$u_1$};   \node[lab] at (-60:\rL) {$\alpha_1$};
    \node[lab] at (-90:\rL) {$v_1$};   \node[lab] at (-120:\rL) {$\beta_1$};
    \node[lab] at (-150:\rL) {$u_2$};  \node[lab] at (180:\rL) {$\alpha_2$};
    \node[lab] at (150:\rL) {$v_2$};   \node[lab] at (120:\rL) {$\beta_2$};
  \end{tikzpicture}%
  \endgroup}

\title{Complexity of liveness in ring Petri nets and of halting in chip-firing on simple digraphs}
\author{Eike Best \and Raymond Devillers \and Petr Jan\v{c}ar \and Mat\v{e}j O\v{s}\v{t}\'{a}dal}
\date{}

\begin{document}
\maketitle

\begin{abstract}
A \emph{ring Petri net} (weighted circuit, cyclic weighted T-system) consists of a single directed circuit
$s_0\,t_0\,s_1\,t_1\cdots s_{n-1}\,t_{n-1}\,s_0$ with arbitrary positive arc weights.
We show that deciding liveness of ring Petri nets is \coNP-complete, and that \coNP-hardness holds in the
strong sense already for \emph{token-conserving} rings, in which every transition consumes and produces the same
number of tokens. The proof is a short reduction from the covering problem for systems of residue classes (the
complement of \textup{\textsc{Simultaneous Incongruences}}). Together with a known reduction to circuits, it
follows that liveness of weighted T-systems (weighted marked graphs, synchronous dataflow graphs) is \coNP-complete,
and \coNP-hard in the strong sense. As a further consequence, the halting problem for chip-firing on simple
directed graphs is \NP-complete, which answers a question of Farrell and Levine. On 7 October 2026, the problem of
liveness was posed to the AI model Claude Opus 5.5 (Anthropic), which found the reduction, also suggesting the
application to the chip-firing problem, and wrote a first version of this paper; the human authors checked it and,
together with the model, revised it iteratively into the present form.
\end{abstract}

\section{Introduction}

Weighted T-systems (also known as weighted marked graphs, weighted event graphs, or synchronous dataflow graphs)
are Petri nets in which every place has exactly one input and one output transition, while arcs may carry
arbitrary weights. For ordinary (plain) T-systems, liveness is decidable in polynomial time: such a system is live
iff every circuit carries a token, or equivalently, iff some firing sequence fires every transition exactly
once~\cite{CHEP71}, \cite[Thm.~5.31]{BD24}. A counterpart of the second criterion is known for weighted
T-systems~\cite{TCCS92}, but there the vector of ones is replaced by a T-invariant that may be exponentially
large~\cite[Sect.~5.10]{BD24}. The \emph{Petri Net Primer}~\cite[Sect.~5.10]{BD24} states:
\begin{quote}\small
``At the date of publishing this book, no polynomial algorithm appears to be known to check the liveness of
weighted T-systems.''
\end{quote}
It includes two weighted T-nets (Figure~5.23 there) for readers wishing to try a ``finger exercise in weighted
T-systems'', even just in weighted cycles, also called ring nets, for which it refers to~\cite{Wei15}. One of the
two nets is taken from~\cite{CR95}, and recent advances in~\cite{Dev23} are mentioned. Liveness of weighted
circuits was related to the Diophantine problem of Frobenius by Chrząstowski-Wachtel and Raczunas~\cite{CR93}.
The complexity of
liveness has also been described as open in the literature on weighted event graphs and synchronous dataflow
graphs. Marchetti and Munier-Kordon~\cite{MMK08} write, about liveness and the maximal throughput of timed
weighted event graphs, that ``the complexity of these two previous problem[s] is still open and is also a
challenging theoretical question'', and a survey talk on synchronous dataflow~\cite{MK22} still lists the
complexity of schedulability (i.e.\ liveness) and of the maximal throughput as unknown; see
also~\cite{MMK09}, where a polynomially computable sufficient condition for liveness is given.

On the upper-bound side, Hujsa, Berthomieu, Dal~Zilio and Le~Botlan showed that a weighted circuit is non-live
iff the state equation has a solution leading to a dead marking~\cite[Thm.~6.6]{HBDL20a}, and even iff it has a
solution leading to an integer vector, possibly with negative entries, at which no transition is
enabled~\cite[Thm.~6.8]{HBDL20a}; with standard bounds on integer programs this puts liveness of ring nets in \coNP{}
for binary-encoded numbers. The same authors established a characterisation of this kind for a class that contains
all strongly connected weighted
T-systems~\cite[Cor.~6.7]{HBDL20b}, and derived a \coNP{} upper bound for the structurally bounded systems of that
class~\cite[Sect.~6.2]{HBDL20b}. This paper supplies a matching lower bound, already for single circuits.

\paragraph{Main result.}
\begin{itemize}[leftmargin=1.4em]
\item \textbf{Theorem~\ref{thm:main}.} Deciding liveness of ring Petri nets is \coNP-complete. Hardness holds
already for token-conserving rings (every transition consumes as many tokens as it produces) whose arc weights and
token numbers are all bounded by a polynomial in the number of places; in particular the problem is \coNP-hard in
the strong sense (i.e.\ also for unary encodings).
\item \textbf{Corollary~\ref{cor:wts}.} Deciding liveness of weighted T-systems is \coNP-complete, and \coNP-hard
in the strong sense.
\end{itemize}
By Lemma~\ref{lem:char}, deadlock-freeness and weak liveness of ring nets coincide with liveness, so they are
\coNP-complete as well. Hardness in Corollary~\ref{cor:wts} holds because ring nets are weighted T-systems, and
membership follows from Proposition~\ref{prop:upper} and a known reduction to circuits~\cite[Prop.~6.3]{HBDL20a}. In
particular, unless $\Ptime=\NP$, liveness of weighted T-systems cannot be decided by a pseudo-polynomial algorithm,
i.e.\ in time polynomial in the size of the system and the largest number occurring in it.

The reduction is from \cover: given residue classes $r_i\bmod b_i$ for $i=0,1,\dots,m-1$, decide
whether every integer lies in one of them. The ring consists of $m$ ``test'' places with output weights $b_i$,
alternating with $m$ ``buffer'' places with output weight $W$, a prime larger than all $b_i$
(Figure~\ref{fig:ring}). The ring behaves like a counter. In each ``round'', every buffer transition fires once;
after $z$ rounds, a further round is possible iff $Wz$ lies in one of the residue classes, and the test place
belonging to the class $r_i\bmod b_i$ checks whether $Wz\equiv r_i\pmod{b_i}$. So the ring deadlocks after
exactly $z^\ast$ rounds, where $z^\ast$ is the least $z\geq0$ such that $Wz$ avoids all residue classes, and it runs
forever if there is no such $z$.

\paragraph{Chip-firing.}
As a consequence, the halting problem for chip-firing on \emph{simple} digraphs is \NP-complete, and \NP-hard
already for strongly connected ones (Section~\ref{sec:chip}). This answers a question of Farrell and
Levine~\cite{FL16}, who proved \NP-completeness for strongly connected multigraphs; their hard instances may have
large edge multiplicities.

Further results, on reversibility and reachability of ring nets and on tractable special cases, are planned for a
later version of this paper.

\paragraph{Organisation.}
Section~\ref{sec:prelim} recalls Petri nets and ring nets. Section~\ref{sec:upper} characterises liveness of ring
nets by deadlock certificates and proves the \coNP{} upper bound. Section~\ref{sec:lower} gives the reduction, the
main result and its extension to weighted T-systems, Section~\ref{sec:chip} treats chip-firing, and
Section~\ref{sec:conclusion} concludes.

\paragraph{How this paper came about.}
On 7 October 2026, the complexity of liveness for ring nets was posed as an open problem to the AI model Claude
Opus 5.5 (Anthropic).
The model found the reduction from \cover{} presented in Section~\ref{sec:lower}, also suggested its application to
the question of Farrell and Levine on chip-firing (Section~\ref{sec:chip}), and produced a first version of this
paper, including computational cross-checks of its claims. The human authors checked the paper (the
statements, the proofs and the use of the literature) and, over several rounds, asked for corrections, more
detailed arguments and further material. These revisions were carried out with the model and led to the present
version, which the authors have checked. The statements of Sections~\ref{sec:upper}--\ref{sec:chip} were also
cross-checked computationally, on exhaustive families of small instances and on many random instances, with two
independent implementations; details are planned for the later version.

\section{Preliminaries}\label{sec:prelim}

\paragraph{Notation.}
We write $\N=\{0,1,2,\dots\}$ for the set of natural numbers (including $0$) and $\Z$ for the set of integers. For
an integer $x$ and an integer $b\geq1$, $\lfloor x/b\rfloor$ is
the largest integer that is at most $x/b$, and
\[
x\bmod b\coloneqq x-b\lfloor x/b\rfloor\in\{0,1,\dots,b-1\}
\]
is the \emph{remainder} of $x$ modulo $b$; this also applies to negative $x$ (for instance, $(-3)\bmod4=1$). We
write $b\mid x$ if $b$ divides $x$, and $x\equiv x'\pmod b$ if $b\mid x-x'$, i.e.\ if $x\bmod b=x'\bmod b$.

Lists are written with their first two members, as in $s_0,s_1,\dots,s_{n-1}$ or $j=1,2,\dots,k$. Such a list
may be shorter than the notation suggests: for $n=1$, the list $s_0,s_1,\dots,s_{n-1}$ consists of $s_0$ only.

For a finite set $X$, a \emph{vector} $v\in\Z^X$ (or $v\in\N^X$) is a function $v\colon X\to\Z$ (or into $\N$).
For $n\in\N$ we also write $n$ for the set $\{0,1,\dots,n-1\}$;
thus the elements of $\Z^n$ are the tuples $v=(v_0,v_1,\dots,v_{n-1})$ of integers. We write $\zero$ for the
\emph{zero vector}, whose entries are all $0$. Vectors are compared componentwise: $u\leq v$ means
$u(x)\leq v(x)$ for all $x\in X$, and $u<v$ means $u(x)<v(x)$ for all $x\in X$. In particular, $v\geq\zero$ means
that all entries of $v$ are non-negative. For a
matrix $C\in\Z^{S\times T}$, i.e.\ a function $C\colon S\times T\to\Z$, and a vector $Y\in\Z^T$, the vector
$C\cdot Y\in\Z^S$ is defined by $(C\cdot Y)(s)=\sum_{t\in T}C(s,t)\,Y(t)$.

\paragraph{Petri nets and markings.}
A \emph{Petri net} (or simply a \emph{net}) is a triple $N=(S,T,F)$ consisting of
\begin{itemize}[leftmargin=1.4em,itemsep=1pt]
\item a finite set $S$ of \emph{places} and a finite set $T$ of \emph{transitions}, with $S\cap T=\emptyset$;
\item a \emph{weight function} $F\colon(S\times T)\cup(T\times S)\to\N$: there is an arc from a place $s$ to a
transition $t$ iff $F(s,t)>0$, and then $F(s,t)$ is the weight of this arc; similarly for arcs from transitions to
places.
\end{itemize}
A \emph{marking} of $N$ is a vector $M\in\N^S$. We think of $M(s)$ as the number of \emph{tokens} on the place
$s$; in particular, markings are non-negative by definition. A \emph{marked net} is a pair $(N,M_0)$ of a net $N$
and a marking $M_0$ of $N$, its \emph{initial marking}. Notions such as the incidence matrix and T-invariants,
introduced below, depend only on $N$; notions such as liveness depend on the marked net. For a transition $t$, the
places $s$ with $F(s,t)>0$ are the \emph{input places} of $t$, and the places $s$ with $F(t,s)>0$ are its
\emph{output places}; similarly, the transitions $t$ with $F(t,s)>0$ and those with $F(s,t)>0$ are the
\emph{input transitions} and the \emph{output transitions} of a place $s$.

\paragraph{Firing.}
A transition $t$ is \emph{enabled} at a marking $M$ if $M(s)\geq F(s,t)$ for all $s\in S$, i.e.\ if every input
place $s$ of $t$ carries at least $F(s,t)$ tokens. An enabled transition $t$ can \emph{fire} (or \emph{occur}).
Firing $t$ removes $F(s,t)$ tokens from every place $s$ and adds $F(t,s)$ tokens to every place $s$; it leads to
the marking $M'$ with
\[
M'(s)=M(s)-F(s,t)+F(t,s)\qquad(s\in S),
\]
and we write $M\fire{t}M'$. (The vector $M'$ is again a marking, i.e.\ non-negative, because $t$ is enabled at
$M$.)

We write $T^\ast$ for the set of finite words over $T$, including the empty word $\varepsilon$. A (finite)
\emph{firing sequence} from a marking $M$ is a word $\sigma=t_1t_2\cdots t_k\in T^\ast$ for which there are
markings $M=M^{(0)},M^{(1)},\dots,M^{(k)}$ with $M^{(j-1)}\fire{t_j}M^{(j)}$ for $j=1,2,\dots,k$; we then write
$M\fire{\sigma}M^{(k)}$, or simply $M\fire{\sigma}$ if the final marking does not matter. The \emph{length} of
$\sigma$ is $|\sigma|=k$. The empty word $\varepsilon$ is a firing sequence of length $0$ from every marking, and
$M\fire{\varepsilon}M$. An \emph{infinite firing sequence} from $M$ is an infinite word $t_1t_2t_3\cdots$ such that
each of its finite prefixes $t_1t_2\cdots t_k$ is a firing sequence from $M$. A marking $M'$ is \emph{reachable}
from $M$ if $M\fire{\sigma}M'$ for some finite firing sequence $\sigma$; the set of all markings reachable from $M$
is denoted by $\reach{M}$ (it contains $M$ itself). The markings in $\reach{M_0}$ are the \emph{reachable markings}
of the marked net $(N,M_0)$.

A marking is \emph{dead} (or a \emph{deadlock}) if it enables no transition. A firing sequence from a marking $M$
is \emph{maximal} if it is infinite, or if it is finite and leads to a dead marking; in other words, if it cannot be
extended by a further transition.

\paragraph{Parikh vectors and the state equation.}
For a finite word $\sigma$ over $T$, the \emph{Parikh vector} $\Par(\sigma)\in\N^T$ counts how often each
transition occurs in $\sigma$: $\Par(\sigma)(t)$ is the number of occurrences of $t$ in $\sigma$; thus
$|\sigma|=\sum_{t\in T}\Par(\sigma)(t)$. The \emph{incidence matrix} of $N$ is the matrix $C\in\Z^{S\times T}$ with
\[
C(s,t)=F(t,s)-F(s,t)\qquad(s\in S,\ t\in T),
\]
the net change of the number of tokens on $s$ caused by one occurrence of $t$. So $M\fire{t}M'$ implies
$M'(s)=M(s)+C(s,t)$ for all $s\in S$. Adding up these changes along a firing sequence gives the \emph{state
equation}: if $M\fire{\sigma}M'$, then
\begin{equation}\label{eq:stateeq}
M'=M+C\cdot\Par(\sigma).
\end{equation}
Formally, this follows by induction on $|\sigma|$: for $\sigma=\varepsilon$ both sides equal $M$, and if
$M\fire{\sigma}M''\fire{t}M'$, then $M'(s)=M''(s)+C(s,t)=M(s)+(C\cdot\Par(\sigma))(s)+C(s,t)=M(s)+(C\cdot\Par(\sigma
t))(s)$ for every $s$. The state equation only records how often each transition occurs, not in which order. Its
converse does not hold in general: a vector $X\in\N^T$ with $M'=M+C\cdot X$ need not be the Parikh vector of any
firing sequence from $M$ to $M'$ (see Example~\ref{ex:running}).

\begin{definition}[liveness notions, after {\cite[Def.~3.10]{BD24}}]\label{def:live}
Let $(N,M_0)$ be a marked net, $N=(S,T,F)$, and let $t\in T$.
\begin{itemize}[leftmargin=1.4em,itemsep=1pt]
\item $t$ is \emph{weakly live} if some infinite firing sequence from $M_0$ contains $t$ infinitely often;
\item $t$ is \emph{(strongly) live} if for every reachable marking $M\in\reach{M_0}$ there is $\sigma\in T^\ast$
with $M\fire{\sigma t}$, i.e.\ if from every reachable marking, $t$ can still occur.
\end{itemize}
The marked net $(N,M_0)$ is \emph{weakly live} or \emph{(strongly) live} if every transition $t\in T$ has the
respective property. It is \emph{deadlock-free} if no reachable marking is dead.
\end{definition}

``Live'' always means strongly live. Spelled out, $(N,M_0)$ is \emph{not} live iff there are a reachable marking
$M$ and a transition $t$ such that $t$ occurs in no firing sequence from $M$, i.e.\ once $M$ has been reached, $t$
can never occur. If $(N,M)$ is live, we also say that $M$ is a \emph{live marking} of $N$, or simply that $M$ is
\emph{live} when $N$ is clear from the context; similarly for the other notions of Definition~\ref{def:live}. Note
that a marking that is not live need not be dead: in Example~\ref{ex:running} below, the marking $(3,0)$ is not
live, although it enables a transition.

\paragraph{Ring nets.}
A \emph{ring net} (or \emph{ring}) with $n\geq1$ places is a net $N=(S,T,F)$ with $S=\{s_0,s_1,\dots,s_{n-1}\}$ and
$T=\{t_0,t_1,\dots,t_{n-1}\}$, where these $2n$ elements are distinct, such that
\[
a_i\coloneqq F(s_i,t_i)\geq1\qquad\text{and}\qquad b_i\coloneqq F(t_i,s_{i+1})\geq1\qquad(0\leq i\leq n-1),
\]
and $F(x,y)=0$ for all other pairs $(x,y)$. Indices are taken modulo $n$ throughout, so that $s_n=s_0$,
$t_{-1}=t_{n-1}$, and so on. So the transition $t_i$ has the single input place $s_i$ and the single output place
$s_{i+1}$, and the places and transitions form the circuit $s_0\,t_0\,s_1\,t_1\cdots s_{n-1}\,t_{n-1}\,s_0$; the
numbers $a_i,b_i$ are the \emph{weights} of the ring. (For $n=1$, the ring consists of the place $s_0$, the
transition $t_0$, an arc from $s_0$ to $t_0$ with weight $a_0$, and an arc from $t_0$ to $s_0$ with weight $b_0$.)
We identify $\Z^T$ with $\Z^n$ via $t_i\mapsto i$, and likewise $\Z^S$ with $\Z^n$ via $s_i\mapsto i$: for
$Y\in\Z^T$ we write $Y_i=Y(t_i)$, and a marking $M$ is written as $(M(s_0),M(s_1),\dots,M(s_{n-1}))$.

The transition $t_i$ is enabled at a marking $M$ iff $M(s_i)\geq a_i$, and firing $t_i$ removes $a_i$ tokens from
$s_i$ and adds $b_i$ tokens to $s_{i+1}$. In particular, a marking $D$ is dead iff $D(s_i)<a_i$ for all $i$. Each
place $s_i$ has exactly one input transition, $t_{i-1}$, and one output transition, $t_i$. Hence the incidence
matrix $C$ of the ring satisfies, for every $Y\in\Z^n$,
\begin{equation}\label{eq:incidence}
(C\cdot Y)(s_i)\;=\;b_{i-1}Y_{i-1}-a_iY_i\qquad(0\leq i\leq n-1):
\end{equation}
place $s_i$ gains $b_{i-1}$ tokens with each occurrence of $t_{i-1}$ and loses $a_i$ tokens with each occurrence of
$t_i$. This also holds for $n=1$, where $t_{i-1}=t_i=t_0$ and $(C\cdot Y)(s_0)=(b_0-a_0)Y_0$.

\begin{figure}[t]
\centering
\begin{tikzpicture}[
  place/.style={circle,draw,thick,minimum size=9mm,inner sep=0pt},
  trans/.style={rectangle,draw,thick,fill=black!15,minimum width=8mm,minimum height=2.4mm,inner sep=0pt},
  arc/.style={-{Latex[length=2.2mm]},thick},
  w/.style={font=\footnotesize,inner sep=2pt}]
  \node[place,label=left:{$s_0$}] (s0) at (0,0) {$4$};
  \node[place,label=right:{$s_1$}] (s1) at (4.4,0) {$0$};
  \node[trans,label=above:{$t_0$}] (t0) at (2.2,1.35) {};
  \node[trans,label=below:{$t_1$}] (t1) at (2.2,-1.35) {};
  \draw[arc] (s0) to[bend left=18] node[w,above left] {$2$} (t0);
  \draw[arc] (t0) to[bend left=18] node[w,above right] {$3$} (s1);
  \draw[arc] (s1) to[bend left=18] node[w,below right] {$6$} (t1);
  \draw[arc] (t1) to[bend left=18] node[w,below left] {$4$} (s0);
\end{tikzpicture}
\caption{The ring of Example~\ref{ex:running}: $n=2$, $(a_0,a_1)=(2,6)$, $(b_0,b_1)=(3,4)$, with the marking
$M_0=(4,0)$. Numbers inside places give the marking, labels at arcs the weights.}
\label{fig:running}
\end{figure}
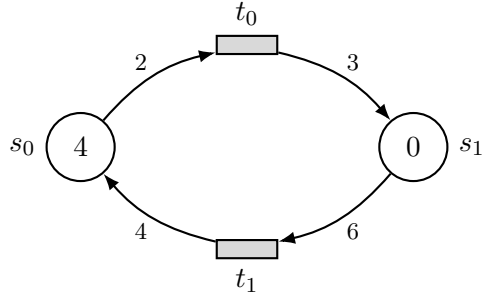

\begin{example}[running example]\label{ex:running}
Let $N$ be the ring with $n=2$, $a_0=2$, $b_0=3$, $a_1=6$ and $b_1=4$ (Figure~\ref{fig:running}): $t_0$ takes $2$
tokens from $s_0$ and puts $3$ tokens on $s_1$, and $t_1$ takes $6$ tokens from $s_1$ and puts $4$ tokens on $s_0$.
Markings are pairs $(M(s_0),M(s_1))$. By~\eqref{eq:incidence},
\[
(C\cdot Y)(s_0)=4Y_1-2Y_0,\qquad (C\cdot Y)(s_1)=3Y_0-6Y_1.
\]
\begin{itemize}[leftmargin=1.4em,itemsep=1pt]
\item From $M_0=(4,0)$ only $t_0$ is enabled, and
$(4,0)\fire{t_0}(2,3)\fire{t_0}(0,6)\fire{t_1}(4,0)$. At $(2,3)$ only $t_0$ is enabled, and at $(0,6)$ only
$t_1$. So the reachable markings of $(N,M_0)$ are $(4,0)$, $(2,3)$ and $(0,6)$, the only maximal firing sequence
from $M_0$ is the infinite sequence $t_0t_0t_1\,t_0t_0t_1\,t_0t_0t_1\cdots$, and $M_0$ is live: from each
reachable marking, both transitions can occur again. The Parikh vector of $t_0t_0t_1$ is $(2,1)$, and indeed
$(4,0)+C\cdot(2,1)=(4+4-4,\;0+6-6)=(4,0)$, in accordance with the state equation~\eqref{eq:stateeq}.
\item From the marking $(3,0)$ we get $(3,0)\fire{t_0}(1,3)$, and $(1,3)$ is dead because $1<2$ and $3<6$. So the
marking $(3,0)$ is not live, although it is not dead. The vector $X=(2,1)$ satisfies $(3,0)+C\cdot X=(3,0)$, but
no firing sequence from $(3,0)$ has the Parikh vector $(2,1)$: the state equation is necessary for reachability,
but not sufficient.
\end{itemize}
\end{example}

\paragraph{T-invariants.}
By the state equation, a firing sequence $\sigma$ from a marking $M$ leads back to $M$ iff
$C\cdot\Par(\sigma)=\zero$. A vector $x\in\N^T$ with $C\cdot x=\zero$ is called a \emph{T-invariant} of $N$; it
describes how often each transition has to occur so that the total effect on the marking is zero. A ring is
\emph{token-conserving} if $a_i=b_i$ for all $i$, i.e.\ if every occurrence of every transition preserves the total
number of tokens.

\begin{lemma}[T-invariants of token-conserving rings]\label{lem:tinv}
Let $N$ be a token-conserving ring with $n$ places and weights $a_i=b_i$, and let
$L\coloneqq\lcm(a_0,a_1,\dots,a_{n-1})$. Then the vector $x_T\in\N^n$ with $x_T(t_i)=L/a_i$ for all $i$ is a
T-invariant of $N$, and the T-invariants of $N$ are exactly the vectors $k\,x_T$ with $k\in\N$. We call $x_T$ the
\emph{minimal T-invariant} of $N$.
\end{lemma}
\begin{proof}
By~\eqref{eq:incidence} with $b_{i-1}=a_{i-1}$, a vector $x\in\N^n$ satisfies $C\cdot x=\zero$ iff
$a_ix_i=a_{i-1}x_{i-1}$ for all $i$, i.e.\ iff the product $a_ix_i$ has the same value $c$ for all $i$. In this case
$x_i=c/a_i$ is an integer for every $i$, so $c\in\N$ is a common multiple of $a_0,a_1,\dots,a_{n-1}$, i.e.\ $c=kL$
for some $k\in\N$, and then $x=k\,x_T$. Conversely, $k\,x_T$ is a T-invariant for every $k\in\N$, since
$a_i\cdot kL/a_i=kL$ for all $i$.
\end{proof}

\paragraph{Encodings and the problem.}
A ring net is given by the number $n$ of places and the weights $a_i,b_i$, and a marking $M$ by the numbers
$M(s_i)$, for $0\leq i\leq n-1$. Unless stated otherwise, numbers are written in binary, and the \emph{size} of an
instance is the total number of bits needed to write it down. A decision problem is in \NP{} if its yes-instances
have certificates of polynomial size that can be checked in polynomial time, and it is in \coNP{} if its complement
is in \NP, i.e.\ if its no-instances have such certificates. A polynomial-time reduction from a problem $A$ to a
problem $B$ is also a reduction from the complement of $A$ to the complement of $B$; hence a problem is
\coNP-hard iff its complement is \NP-hard. A problem is \coNP-hard \emph{in the strong sense} if it remains
\coNP-hard when restricted to instances whose numbers are bounded by a polynomial in the size of the instance;
equivalently, if it is \coNP-hard even when all numbers are written in unary; \NP-hardness in the strong sense is
defined in the same way. We study the problem
\begin{quote}
\pb: given a ring net $N$ and a marking $M_0$ of $N$, is $M_0$ live, i.e.\ is the marked ring $(N,M_0)$ live?
\end{quote}

\section{Deadlock certificates and the upper bound}\label{sec:upper}

The results of this section are essentially known, from the work of Hujsa, Berthomieu, Dal~Zilio and Le~Botlan:
for rings, the equivalence (1)$\Leftrightarrow$(6) in Lemma~\ref{lem:char} is \cite[Thm.~6.6]{HBDL20a}, the
equivalence (1)$\Leftrightarrow$(5) is \cite[Thm.~6.8]{HBDL20a}, and Corollary~\ref{cor:unique} is close
to~\cite[Thm.~6.7]{HBDL20a}. The \coNP{} upper bound (Proposition~\ref{prop:upper}) follows from
(1)$\Leftrightarrow$(6) directly with standard bounds on integer programs, and the same authors stated such a bound
for the structurally bounded systems of a larger class~\cite[Sect.~6.2]{HBDL20b}.
We give short direct proofs for rings, because Sections~\ref{sec:lower} and~\ref{sec:chip} use these results in
exactly this form.

Throughout this section, $N$ is a ring net with the $n$ places $s_0,s_1,\dots,s_{n-1}$, the transitions
$t_0,t_1,\dots,t_{n-1}$, the weights $a_i,b_i$ and the incidence matrix $C$, and $M_0$ is a marking of $N$; indices
are taken modulo $n$. For a finite firing sequence $\sigma$ we write $\Par(\sigma)_i$ for the number of occurrences of
$t_i$ in $\sigma$.

\paragraph{The idea.}
If $M_0$ is not live, then some firing sequence $\rho$ leads from $M_0$ to a dead marking $D$ (this is shown in
Lemma~\ref{lem:char} below). By the state equation, its Parikh vector $Y=\Par(\rho)$ satisfies $D=M_0+C\cdot Y$, so
$Y$ witnesses that $M_0$ is not live. The main point of this section (Lemma~\ref{lem:bound}) is a converse: any
vector $Y\in\N^n$ for which $M_0+C\cdot Y$ is below all enabling thresholds bounds all behaviours of $(N,M_0)$, even
if $Y$ is not the Parikh vector of any firing sequence. So non-liveness has certificates that are easy to check, and
it remains to show that small certificates exist.

\paragraph{Deadlock certificates.}
A vector $Y\in\N^n$ is a \emph{deadlock certificate} for $(N,M_0)$ if
\begin{equation}\label{eq:cert}
(M_0+C\cdot Y)(s_i)<a_i\qquad\text{for all }i\in\{0,1,\dots,n-1\}.
\end{equation}
We do not require $M_0+C\cdot Y\geq\zero$. If $M_0+C\cdot Y\geq\zero$ does hold, then $M_0+C\cdot Y$ is a marking
at which no $t_i$ is enabled, i.e.\ a dead marking. In particular, by the state equation, the Parikh vector of every
firing sequence from $M_0$ to a dead marking is a deadlock certificate. In Example~\ref{ex:running} with the
marking $M_0=(3,0)$, the vector $Y=(1,0)$ is a deadlock certificate: $(3,0)+C\cdot Y=(1,3)$, and $1<2$, $3<6$.

\begin{lemma}[certificates bound all behaviours]\label{lem:bound}
Let $Y$ be a deadlock certificate for $(N,M_0)$. Then $\Par(\sigma)\leq Y$ for every finite firing sequence
$\sigma$ from $M_0$. Consequently, every firing sequence from $M_0$ has length at most $\sum_iY_i$, and there is no
infinite firing sequence from $M_0$.
\end{lemma}

\noindent\emph{Idea.} Consider the first moment at which some transition $t_i$ would occur for the $(Y_i+1)$-st
time. Up to this moment, $t_{i-1}$ has occurred at most $Y_{i-1}$ times, so $s_i$ has received at most
$b_{i-1}Y_{i-1}$ tokens, and after $Y_i$ occurrences of $t_i$ fewer than $a_i$ tokens are left on $s_i$,
by~\eqref{eq:cert}. So $t_i$ cannot occur once more. The proof makes this precise by induction.

\begin{proof}
We show $\Par(\sigma)\leq Y$ by induction on $|\sigma|$. For $\sigma=\varepsilon$, we have
$\Par(\varepsilon)=\zero\leq Y$, since $Y\in\N^n$.

For the induction step, let $\sigma t_i$ be a firing sequence from $M_0$, say $M_0\fire{\sigma}M\fire{t_i}$, and
assume $\Par(\sigma)\leq Y$. The vector $\Par(\sigma t_i)$ differs from $\Par(\sigma)$ only in its $i$-th entry,
which is larger by $1$. If $\Par(\sigma)_i<Y_i$, then $\Par(\sigma t_i)_i\leq Y_i$, and we are done. By the
induction hypothesis, the only other case is $\Par(\sigma)_i=Y_i$; we show that it cannot occur. By the state
equation~\eqref{eq:stateeq} and by~\eqref{eq:incidence}, the number of tokens on $s_i$ at $M$ is
\[
M(s_i)=M_0(s_i)+b_{i-1}\Par(\sigma)_{i-1}-a_i\Par(\sigma)_i
\;\leq\;M_0(s_i)+b_{i-1}Y_{i-1}-a_iY_i=(M_0+C\cdot Y)(s_i)<a_i,
\]
where we used $\Par(\sigma)_{i-1}\leq Y_{i-1}$ (induction hypothesis), $b_{i-1}>0$, $\Par(\sigma)_i=Y_i$
and~\eqref{eq:cert}. But $t_i$ is enabled at $M$, i.e.\ $M(s_i)\geq a_i$, a contradiction. (For $n=1$, the indices
$i-1$ and $i$ coincide, $\Par(\sigma)_{i-1}=\Par(\sigma)_i=Y_i$, and the same computation applies.)

Finally, $|\sigma|=\sum_i\Par(\sigma)_i\leq\sum_iY_i$ for every finite firing sequence $\sigma$ from $M_0$. An
infinite firing sequence from $M_0$ would have finite prefixes of every length, which is impossible.
\end{proof}

\begin{lemma}[a transition that stops firing stops the whole ring]\label{lem:propagate}
Let $M$ be a marking of $N$.
\begin{enumerate}[label=(\alph*),itemsep=1pt]
\item In every infinite firing sequence from $M$, every transition occurs infinitely often.
\item If some transition occurs in no firing sequence from $M$, then a dead marking is reachable from $M$.
\end{enumerate}
\end{lemma}

\noindent\emph{Idea.} Tokens arrive at $s_i$ only through $t_{i-1}$. So if one transition stops firing for good,
then so does the next one, then the one after it, and so on around the ring, until no transition can fire any more.

\begin{proof}
(a) Let $\tau_1\tau_2\tau_3\cdots$ be an infinite firing sequence from $M$. Since there are only $n$ transitions,
some transition occurs in it infinitely often. Suppose, for a contradiction, that some transition occurs in it only
finitely often. Starting at a transition that occurs infinitely often and going backwards around the ring, i.e.\
through $t_j,t_{j-1},t_{j-2},\dots$, we eventually meet a transition that occurs only finitely often. Hence there is
an index $i$ such that $t_i$ occurs infinitely often, but $t_{i-1}$ only finitely often. (This is impossible for
$n=1$, where $t_{i-1}=t_i$; so for $n=1$ there is nothing more to show.) Choose $k$ such that $t_{i-1}$ does not
occur among $\tau_{k+1},\tau_{k+2},\dots$, and let $M\fire{\tau_1\tau_2\cdots\tau_k}M'$. After position $k$, no
tokens are added to $s_i$, because $t_{i-1}$ is the only input transition of $s_i$, and each occurrence of $t_i$
removes $a_i\geq1$ tokens from $s_i$. Since the number of tokens on $s_i$ never becomes negative, $t_i$ occurs at
most $M'(s_i)/a_i$ times after position $k$. This contradicts the choice of $i$.

(b) Let $t$ be a transition that occurs in no firing sequence from $M$, and suppose, for a contradiction, that no
dead marking is reachable from $M$. Then every marking reachable from $M$ enables some transition. Starting at $M$
and firing, at each step, some transition that is enabled at the current marking (which is again reachable from
$M$), we obtain an infinite firing sequence from $M$. By~(a), $t$ occurs in it, say at position $k$. Then the
prefix of length $k$ is a firing sequence from $M$ in which $t$ occurs, a contradiction.
\end{proof}

\begin{lemma}[characterisation]\label{lem:char}
For a ring net $N$ and a marking $M_0$ of $N$ the following are equivalent:
\begin{enumerate}[label=(\arabic*),itemsep=1pt]
\item $(N,M_0)$ is live;
\item $(N,M_0)$ is deadlock-free;
\item $(N,M_0)$ is weakly live;
\item there is an infinite firing sequence from $M_0$;
\item $(N,M_0)$ has no deadlock certificate, i.e.\ there is no $Y\in\N^n$ with $(M_0+C\cdot Y)(s_i)<a_i$ for
all $i$;
\item there is no $Y\in\N^n$ such that $M_0+C\cdot Y$ is a dead marking, i.e.\
$0\leq(M_0+C\cdot Y)(s_i)<a_i$ for all~$i$.
\end{enumerate}
\end{lemma}
\begin{proof}
(1)$\Rightarrow$(2): Suppose that some reachable marking $D$ is dead. The only firing sequence from $D$ is the
empty one, so $t_0$ occurs in no firing sequence from $D$, and $t_0$ is not live. Hence $(N,M_0)$ is not live.

(2)$\Rightarrow$(4): Suppose that no reachable marking is dead. Since $M_0$ is reachable, some transition $\tau_1$
is enabled at $M_0$; let $M_0\fire{\tau_1}M^{(1)}$. The marking $M^{(1)}$ is reachable, hence not dead, so some
transition $\tau_2$ is enabled at $M^{(1)}$; let $M^{(1)}\fire{\tau_2}M^{(2)}$; and so on. This yields an infinite
firing sequence $\tau_1\tau_2\tau_3\cdots$ from $M_0$.

(4)$\Rightarrow$(5): If $(N,M_0)$ had a deadlock certificate, then by Lemma~\ref{lem:bound} there would be no
infinite firing sequence from $M_0$.

(5)$\Rightarrow$(6): A vector $Y$ as in (6) satisfies~\eqref{eq:cert}, i.e.\ it is a deadlock certificate. So if
there is no deadlock certificate, there is no vector as in (6) either.

(6)$\Rightarrow$(1): We show the contrapositive. Suppose that $(N,M_0)$ is not live. Then there are a reachable
marking $M$ and a transition $t_i$ such that $t_i$ occurs in no firing sequence from $M$. By
Lemma~\ref{lem:propagate}(b), a dead marking $D$ is reachable from $M$, and hence from $M_0$; say
$M_0\fire{\rho}D$. By the state equation, $Y\coloneqq\Par(\rho)\in\N^n$ satisfies $M_0+C\cdot Y=D$, which is a dead
marking. So (6) fails.

(3)$\Rightarrow$(4): Every transition is weakly live, so in particular some infinite firing sequence from $M_0$
exists (recall $n\geq1$).

(4)$\Rightarrow$(3): By Lemma~\ref{lem:propagate}(a), every transition occurs infinitely often in an infinite
firing sequence from $M_0$; so every transition is weakly live.
\end{proof}

As mentioned at the beginning of this section, the equivalences (1)$\Leftrightarrow$(6) and
(1)$\Leftrightarrow$(5) are \cite[Thm.~6.6 and Thm.~6.8]{HBDL20a}; see~\cite[Cor.~6.7]{HBDL20b} for a
generalisation.

Conditions (5) and (6) differ only in whether $M_0+C\cdot Y\geq\zero$ is required. Condition (6) speaks about
vectors $Y$ that lead, via the state equation, to an actual dead marking; condition (5) only asks that every place
$s_i$ carries fewer than $a_i$ ``tokens'', and allows negative entries. The implication (5)$\Rightarrow$(6) is
trivial; the converse holds by the lemma, but the proof goes through liveness. The proof of the upper bound below
(Proposition~\ref{prop:upper}) uses the form~(6).

Lemmas~\ref{lem:bound} and~\ref{lem:char} together show that a marking that is not live has essentially only one
way of dying:

\begin{corollary}[the dead marking is unique; cf.\ {\cite[Thm.~6.7]{HBDL20a}}]\label{cor:unique}
Suppose that $M_0$ is not live. Then $(N,M_0)$ has a least deadlock certificate $Y_{\min}$, i.e.\ a deadlock
certificate with $Y_{\min}\leq Y$ for every deadlock certificate $Y$. Every maximal firing sequence from $M_0$ is
finite, has the Parikh vector $Y_{\min}$, and ends in the dead marking $M_0+C\cdot Y_{\min}$; in particular, this is
the only dead marking reachable from $M_0$.
\end{corollary}

\noindent\emph{Idea.} By Lemma~\ref{lem:bound}, the Parikh vector of every firing sequence from $M_0$ is bounded
by every deadlock certificate. The Parikh vector of a maximal firing sequence is itself a deadlock certificate, so it
is the least one, and therefore the same for all maximal firing sequences.

\begin{proof}
Let $\rho$ be a maximal firing sequence from $M_0$ (one always exists: fire enabled transitions as long as
possible). Since $M_0$ is not live, there is no infinite firing sequence from $M_0$, by Lemma~\ref{lem:char},
(4)$\Rightarrow$(1); so $\rho$, like every maximal firing sequence from $M_0$, is finite and leads to a dead
marking, say $M_0\fire{\rho}D$. By the state equation, $D=M_0+C\cdot\Par(\rho)$, so $\Par(\rho)$ is a deadlock certificate, and $\Par(\rho)\leq Y$ for
every deadlock certificate $Y$, by Lemma~\ref{lem:bound}. If $\rho'$ is another maximal firing sequence from $M_0$,
then likewise $\Par(\rho')\leq\Par(\rho)$ and $\Par(\rho)\leq\Par(\rho')$, i.e.\ $\Par(\rho')=\Par(\rho)$. So
$Y_{\min}\coloneqq\Par(\rho)$ has all the stated properties. Finally, every dead marking reachable from $M_0$ is
reached by a maximal firing sequence.
\end{proof}

So, whatever choices are made, a ring started at a marking that is not live always stops in the same dead marking,
after the same numbers of occurrences of each transition. Chip-firing games have the same property
(Section~\ref{sec:chip}).

\begin{proposition}[upper bound; cf.\ {\cite{HBDL20a,HBDL20b}}]\label{prop:upper}
\pb{} is in \coNP.
\end{proposition}
\begin{proof}
We show that the complement of \pb{} is in \NP. By Lemma~\ref{lem:char}, (1)$\Leftrightarrow$(6), $M_0$ is not
live iff there is $Y\in\N^n$ such that $D\coloneqq M_0+C\cdot Y$ satisfies $0\leq D(s_i)\leq a_i-1$ for all $i$.
With a vector $Z\in\N^n$ of slack variables, this is the existence of a solution of the integer linear system
\[
D-C\cdot Y=M_0,\qquad D+Z=(a_0-1,a_1-1,\dots,a_{n-1}-1),\qquad D,Y,Z\in\N^n,
\]
a system of the form $A\,v=c$, $v\in\N^q$, with $p=2n$ equations and $q=3n$ unknowns $v=(D,Y,Z)$. The entries
of the $p\times q$ matrix $A$ are $0$, $\pm1$ and the entries of $-C$, and the entries of $c$ are the numbers
$M_0(s_i)$ and $a_i-1$; let $w$ be the largest absolute value of all these numbers. By the theorem of
Papadimitriou~\cite[Theorem, p.~767]{Pap81}, if a system $A\,v=c$ with $p$ equations and $q$ unknowns, all of whose
coefficients have absolute value at most $w$, has a solution $v\in\N^q$, then it has one with all entries at most
$q\,(p\,w)^{2p+1}$. This bound has $O(n\log(nw))$ bits, which is polynomial in the size of the instance, because
$\log w$ is at most the number of bits of the largest number in it. Hence $M_0$ is not live iff there is a vector
$Y\in\N^n$ whose entries have polynomially many bits such that $\zero\leq M_0+C\cdot Y<(a_0,a_1,\dots,a_{n-1})$. A
nondeterministic algorithm can guess such a $Y$ and check the condition in polynomial time, by computing the $n$
numbers $(M_0+C\cdot Y)(s_i)$ using~\eqref{eq:incidence}. So the complement of \pb{} is in \NP, i.e.\ \pb{} is in
\coNP.
\end{proof}

Bounds of this kind on small nonnegative integer solutions of linear systems go back to Borosh and
Treybig~\cite{BT76} and von zur Gathen and Sieveking~\cite{GS78}; we use Papadimitriou's form because it is stated
for equations $A\,v=c$ with $v\geq\zero$ and gives an explicit bound.

\begin{example}[running example, continued]\label{ex:running2}
In Example~\ref{ex:running}, $(C\cdot Y)(s_0)=-2(Y_0-2Y_1)$ and $(C\cdot Y)(s_1)=3(Y_0-2Y_1)$; so $C\cdot Y$
depends only on $d\coloneqq Y_0-2Y_1$. Consider the marking $M_0=(c,0)$ with $c\in\N$. A vector $Y\in\N^2$ is a
deadlock certificate for $(N,M_0)$ iff
\[
c-2d<2\qquad\text{and}\qquad3d<6,\qquad\text{i.e.}\qquad \frac{c-2}{2}<d\leq1 .
\]
For $c\geq4$ no integer $d$ satisfies this, so $(c,0)$ is live by Lemma~\ref{lem:char}, as observed for $c=4$ in
Example~\ref{ex:running}. For $c\leq3$, the value $d=1$ satisfies it: every $Y\in\N^2$ with $Y_0-2Y_1=1$, for
instance $Y=(1,0)$ or $Y=(3,1)$, is a deadlock certificate, and $(c,0)$ is not live. For such $Y$,
$M_0+C\cdot Y=(c-2,3)$. For $c\in\{2,3\}$ this is a dead marking; for $c=3$ it is the dead marking $(1,3)$ found in
Example~\ref{ex:running}, and Lemma~\ref{lem:bound} with $Y=(1,0)$ confirms that from $(3,0)$, $t_0$ can occur at
most once and $t_1$ not at all. For $c\in\{0,1\}$, the vector $(c-2,3)$ has a negative entry, so $Y=(1,0)$ is a
deadlock certificate in the sense of (5) but does not satisfy (6); for these $c$, the vector $Y=(0,0)$ satisfies
(6), since $(c,0)$ is dead itself. In accordance with Corollary~\ref{cor:unique}, the least deadlock certificate is
$Y_{\min}=(1,0)$ for $c\in\{2,3\}$, the Parikh vector of the only maximal firing sequence~$t_0$, and
$Y_{\min}=(0,0)$ for $c\in\{0,1\}$, where the only maximal firing sequence is the empty one.
\end{example}

\begin{remark}[more tokens never hurt]\label{rem:upward}
Firing is \emph{monotone}~\cite[Lemma~1.13]{BD24}: if $M\fire{\sigma}M'$ and $K$ is any marking, then
$M+K\fire{\sigma}M'+K$, because additional tokens never prevent a transition from being enabled and are not touched
when it fires; the same holds for infinite firing sequences, prefix by prefix. By Lemma~\ref{lem:char}, a marking of
a ring is live iff some infinite firing sequence starts at it. Hence, if $M$ is a live marking of a ring $N$, then
every marking $M'\geq M$ of $N$ is live as well. So the set of live markings of a ring is \emph{upward closed}; by
Dickson's lemma, it has only finitely many minimal elements, and it consists of the markings that are greater than
or equal to one of them. In general Petri nets, in contrast, adding tokens can destroy
liveness~\cite[Sect.~5.1 and Exercise~5.5]{BD24}.
\end{remark}

\section{The lower bound}\label{sec:lower}

In this section we show that \pb{} is \coNP-hard, even in the strong sense. The reduction starts from the covering
problem for residue classes (Definition~\ref{def:covering}), which is \coNP-complete
(Proposition~\ref{prop:cover}). For a covering instance $I$ we construct a token-conserving marked ring
$R(I,W)$ (Construction~\ref{con:ring}) and show that it is live iff $I$ covers $\Z$ (Lemma~\ref{lem:key}).
Example~\ref{ex:twoinstances} illustrates the construction on two small instances, and Theorem~\ref{thm:main} puts
the pieces together. Finally, Corollary~\ref{cor:wts} extends the result to weighted T-systems.

\begin{definition}[covering]\label{def:covering}
A \emph{covering instance} is a list $I=((r_0,b_0),(r_1,b_1),\dots,(r_{m-1},b_{m-1}))$ of $m\geq1$ pairs of
integers with $b_i\geq1$ and $0\leq r_i<b_i$. An integer $x$ is \emph{covered} by $I$ if $x\equiv r_i\pmod{b_i}$ for some $i$; $I$
\emph{covers} $\Z$ if every integer is covered. \cover{} asks whether a given $I$ covers~$\Z$.
\end{definition}

For instance, $((0,2),\allowbreak(1,2))$ covers $\Z$ (every integer is even or odd), and so does
$((0,2),\allowbreak(1,4),\allowbreak(3,4))$: an odd integer leaves the remainder $1$ or $3$ when divided by $4$. In
contrast, $((0,2),\allowbreak(1,4))$ does not cover $\Z$,
since the integers $x$ with $x\bmod4=3$ are not covered. A less obvious covering, due to Erdős~\cite{Erd50},
consists of the classes $0\bmod2$, $0\bmod3$, $1\bmod4$, $5\bmod6$ and $7\bmod12$.

For a covering instance $I$ as above, let $B\coloneqq\lcm(b_0,b_1,\dots,b_{m-1})$. Since every $b_i$ divides $B$, two
integers $x\equiv x'\pmod B$ satisfy $x\equiv x'\pmod{b_i}$ for every $i$, so $x$ is covered by $I$ iff $x'$ is.
In particular,
\begin{equation}\label{eq:period}
I\text{ covers }\Z\quad\iff\quad\text{every }x\in\{0,1,\dots,B-1\}\text{ is covered by }I.
\end{equation}

The complement of \cover{} is \textup{\textsc{Simultaneous Incongruences}}~\cite[AN2]{GJ79}. The idea of
encoding truth assignments by residues modulo distinct primes goes back to Stockmeyer and Meyer~\cite{SM73}, who
used it to show that universality of regular expressions over a one-letter alphabet is \coNP-complete; the residue
class $r\bmod b$ corresponds to the expression $0^r(0^b)^\ast$. Problems about residue classes of this kind also
underlie the (strong) \coNP-hardness of deciding the feasibility of periodic real-time task systems with release
offsets~\cite{LM80,BRH90}. We give a self-contained proof for residue classes, which also records that the moduli
stay small.

Recall that a \emph{literal} is a Boolean variable $X$ or its negation $\neg X$, a \emph{clause} is a disjunction
of literals, and a \emph{3-CNF formula} is a conjunction of clauses, each consisting of three literals.

\begin{proposition}[cf.\ {\cite{SM73}}, {\cite[AN2]{GJ79}}]\label{prop:cover}
\cover{} is \coNP-complete. More precisely, let $p_j$ denote the $j$-th odd prime ($p_1=3$, $p_2=5$,
$p_3=7,\dots$). There is a polynomial-time reduction mapping each 3-CNF formula $\varphi$ that has at least one
clause and whose clauses consist of three literals over three distinct variables, with $k$ variables and $l$
clauses, to a covering instance
$I(\varphi)$ with $m=l+\sum_{j=1}^{k}(p_j-2)$ pairs, all of whose moduli are at most
$p_k^3=O(k^3\log^3k)$, such that $\varphi$ is unsatisfiable iff $I(\varphi)$ covers~$\Z$.
\end{proposition}
\begin{proof}
\emph{Membership in \coNP.} If $I$ does not cover $\Z$, then by~\eqref{eq:period} some
$x\in\{0,1,\dots,B-1\}$ is uncovered. Since $B$ is at most the product $b_0b_1\cdots b_{m-1}$, the binary
representation of $x$ is not longer than the binary representations of $b_0,b_1,\dots,b_{m-1}$ together, and one
can check in polynomial time that
$x\not\equiv r_i\pmod{b_i}$ for all $i$. So such an $x$ is a certificate of polynomial size for non-covering.

\emph{Normal form of the formulas.} 3-SAT remains \NP-complete for formulas that have at least one clause and
whose clauses consist of three literals over three distinct variables. Indeed, a clause containing a variable
together with its negation is satisfied by every assignment and can be deleted, and a clause with fewer than three
distinct literals can be padded with new variables without changing satisfiability: $(\ell_1\vee\ell_2)$ is
replaced by $(\ell_1\vee\ell_2\vee Z)\wedge(\ell_1\vee\ell_2\vee\neg Z)$, and $(\ell_1)$ by the four clauses
$(\ell_1\vee\pm Z_1\vee\pm Z_2)$, where $Z,Z_1,Z_2$ are new variables. A formula without clauses is satisfiable and
can be replaced by the single clause $(Z_1\vee Z_2\vee Z_3)$ over three new variables. In particular, such a formula
has $k\geq3$ variables, and the instance $I(\varphi)$ constructed below has $m\geq1$ pairs. Finally, repeated clauses
can be deleted, so 3-SAT remains \NP-complete also for formulas of this form without repeated clauses.

\emph{The reduction.} Let $\varphi$ be such a formula. Its variables are $X_1,X_2,\dots,X_k$, and its clauses
are $c_1,c_2,\dots,c_l$. For a clause $c$ over the variables $X_{j_1},X_{j_2},X_{j_3}$ (with $j_1,j_2,j_3$ distinct), let
$(f^c_1,f^c_2,f^c_3)\in\{0,1\}^3$ be the unique assignment to these three variables that falsifies $c$, writing
$1$ for true and $0$ for false: $f^c_t=0$ if $X_{j_t}$ occurs positively in $c$, and $f^c_t=1$ if it occurs
negatively. Then $I(\varphi)$ consists of the following pairs:
\begin{enumerate}[label=(\roman*),leftmargin=2em,itemsep=1pt]
\item the pairs $(\rho,p_j)$ for $1\leq j\leq k$ and $2\leq\rho\leq p_j-1$;
\item for each clause $c$ as above, the pair $(\rho_c,q_c)$, where $q_c\coloneqq p_{j_1}p_{j_2}p_{j_3}$ and
$\rho_c$ is the unique integer in $[0,q_c)$ with $\rho_c\equiv f^c_t\pmod{p_{j_t}}$ for $t=1,2,3$; it exists by
the Chinese remainder theorem, since $p_{j_1},p_{j_2},p_{j_3}$ are distinct primes.
\end{enumerate}

\emph{Size and running time.} Group~(i) has $p_j-2\geq1$ pairs for each $j$, and group~(ii) has $l$ pairs, so
$m=l+\sum_{j=1}^k(p_j-2)$. All moduli are at most $p_k^3$, and $p_k=O(k\log k)$ by the prime number theorem.
The primes $p_1,p_2,\dots,p_k$ can be found by trial division, and each $\rho_c$ by the
extended Euclidean algorithm, so $I(\varphi)$ can be computed in time polynomial in $k+l$.

\emph{Correctness.} We use the following consequence of the Chinese remainder theorem: for every integer $x$ and
every clause $c$ over $X_{j_1},X_{j_2},X_{j_3}$,
\begin{equation}\label{eq:crt}
x\equiv\rho_c\pmod{q_c}\quad\iff\quad x\bmod p_{j_t}=f^c_t\ \text{ for }t=1,2,3.
\end{equation}
Indeed, since $q_c$ is the product of the distinct primes $p_{j_1},p_{j_2},p_{j_3}$, we have
$x\equiv\rho_c\pmod{q_c}$ iff $x\equiv\rho_c\equiv f^c_t\pmod{p_{j_t}}$ for $t=1,2,3$, and $0\leq f^c_t<p_{j_t}$.

Suppose first that some integer $x$ is not covered by $I(\varphi)$. Since $x$ is not covered by the pairs~(i),
$x\bmod p_j<2$, i.e.\ $x\bmod p_j\in\{0,1\}$, for every $j$. Let $A$ be the assignment with
$A(X_j)\coloneqq x\bmod p_j$. Since $x$ is not covered by the pairs~(ii), \eqref{eq:crt} shows that for every
clause $c$ the values of $A$ on the variables of $c$ differ from $(f^c_1,f^c_2,f^c_3)$, so $A$ does not falsify
$c$. Hence $A$ satisfies $\varphi$.

Conversely, let $A$ be a satisfying assignment, with values in $\{0,1\}$. By the Chinese remainder theorem there
is an integer $x$ with $x\equiv A(X_j)\pmod{p_j}$ for all $j$. Then $x\bmod p_j=A(X_j)\in\{0,1\}$, so $x$ is not
covered by the pairs~(i); and since $A$ falsifies no clause, \eqref{eq:crt} shows that $x$ is not covered by the
pairs~(ii).

Thus $\varphi$ is satisfiable iff some integer is not covered by $I(\varphi)$, i.e.\ $\varphi$ is unsatisfiable
iff $I(\varphi)$ covers $\Z$. Since 3-SAT is \NP-complete, \cover{} is \coNP-hard.
\end{proof}

\paragraph{From covering instances to rings.}
We now come to the core of the reduction. By Proposition~\ref{prop:cover}, it suffices to transform each covering
instance $I=((r_0,b_0),(r_1,b_1),\dots,(r_{m-1},b_{m-1}))$, in polynomial time and with polynomially bounded
numbers, into a marked ring that is live iff $I$ covers~$\Z$. Recall from Section~\ref{sec:prelim} that in a ring,
the transition $t_i$ takes $a_i$ tokens from $s_i$ and puts $b_i$ tokens on $s_{i+1}$, and that the ring is
token-conserving if $a_i=b_i$ for all $i$; by Lemma~\ref{lem:char}, a marked ring is live iff it has an infinite
firing sequence. The ring $R(I,W)$ constructed below is token-conserving; its weights are the moduli of $I$ and a
number $W$ that is larger than all moduli and coprime to them. It has one pair of places $u_i,v_i$ for each pair
$(r_i,b_i)$ of $I$. Together, these pairs act as a counter $z$ that can advance from $z$ to $z+1$ iff $Wz$ is
covered by $I$, as explained after the construction. So the ring runs forever iff all the numbers
$0,W,2W,3W,\dots$ are covered, and because $W$ is coprime to all moduli, this holds iff $I$ covers~$\Z$.

\begin{construction}[the marked ring $R(I,W)$]\label{con:ring}
Let $I=((r_0,b_0),(r_1,b_1),\dots,(r_{m-1},b_{m-1}))$ be a covering instance and let $W$ be an integer with
$W>\max_ib_i$ and $\gcd(W,b_i)=1$ for all $i$ (for instance, a prime $W>\max_ib_i$). Put
$\bar r_i\coloneqq(-r_i)\bmod b_i$; thus $0\leq\bar r_i<b_i$ and $\bar r_i\equiv-r_i\pmod{b_i}$. The underlying
ring net has the $2m$ places $u_0,v_0,u_1,v_1,\dots,u_{m-1},v_{m-1}$, in this cyclic order, and the set $T$ of $2m$
transitions $\alpha_0,\beta_0,\alpha_1,\beta_1,\dots,\alpha_{m-1},\beta_{m-1}$:
\begin{itemize}[leftmargin=1.4em,itemsep=1pt]
\item $\alpha_i$ takes $b_i$ tokens from $u_i$ and puts $b_i$ tokens on $v_i$;
\item $\beta_i$ takes $W$ tokens from $v_i$ and puts $W$ tokens on $u_{i+1}$ (where $u_m=u_0$).
\end{itemize}
Its initial marking $M_0$ is given by $M_0(u_i)=\bar r_i$ and $M_0(v_i)=W-\bar r_i$; both are non-negative, as
$\bar r_i<b_i<W$. The marked ring $R(I,W)$ is the pair of this ring net and $M_0$; so ``$R(I,W)$ is live'' means
that $M_0$ is live. The ring is token-conserving, and $M_0$ puts $mW$ tokens on it in total. (In the notation of
Section~\ref{sec:prelim}, with $s_{2i}=u_i$, $s_{2i+1}=v_i$, $t_{2i}=\alpha_i$ and $t_{2i+1}=\beta_i$, both arc
weights of $\alpha_i$ are $b_i$ and both arc weights of $\beta_i$ are $W$.) Indices of $u_i,v_i,\alpha_i,\beta_i$
are taken modulo $m$, as for rings in Section~\ref{sec:prelim}; for instance, $\beta_{-1}=\beta_{m-1}$ is the input
transition of $u_0$.
\end{construction}

\begin{figure}[t]
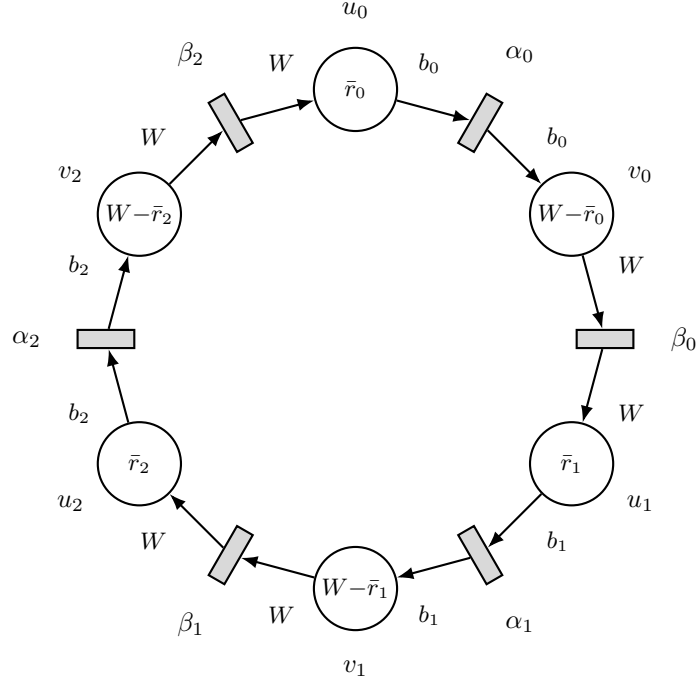

\centering
\ringRIW{}
\caption{The ring $R(I,W)$ for $m=3$. Numbers inside places give the initial marking, labels at arcs the
weights; $\bar r_i=(-r_i)\bmod b_i$, and $W>\max_ib_i$ is coprime to all $b_i$. Tokens flow clockwise.}
\label{fig:ring}
\end{figure}

\paragraph{How the ring works.}
Look at one pair of places $u_i,v_i$. Initially it holds $W$ tokens. Each occurrence of $\beta_{i-1}$ brings $W$
more tokens to $u_i$, each occurrence of $\beta_i$ takes $W$ tokens from $v_i$, and $\alpha_i$ only moves tokens
inside the pair, from $u_i$ to $v_i$, $b_i$ at a time. Suppose that $\beta_{i-1}$ and $\beta_i$ have both occurred
$z$ times, and $\alpha_i$ as often as it could. Then the pair holds $W$ tokens again. The place $u_i$ has received
$\bar r_i+Wz$ tokens in total and has passed them on in groups of $b_i$, so it holds the remainder
$(\bar r_i+Wz)\bmod b_i$, and $v_i$ holds the other $W-(\bar r_i+Wz)\bmod b_i$ tokens. Hence $\beta_i$ can occur for
the $(z+1)$-st time iff this remainder is $0$, i.e.\ iff $Wz\equiv r_i\pmod{b_i}$: the pair $u_i,v_i$ tests
whether $Wz$ lies in the residue class $r_i\bmod b_i$ of~$I$. So the common number $z$ of occurrences of the transitions $\beta_j$
acts as a counter, which can advance from $z$ to $z+1$ iff $Wz$ is covered by $I$. Lemma~\ref{lem:key} below makes
this precise.

Let $C$ be the incidence matrix of $R(I,W)$. Place $u_i$ gains $W$ tokens with each occurrence of $\beta_{i-1}$
and loses $b_i$ tokens with each occurrence of $\alpha_i$; place $v_i$ gains $b_i$ tokens with each occurrence of
$\alpha_i$ and loses $W$ tokens with each occurrence of $\beta_i$. So, by~\eqref{eq:incidence}, for every
$Y\in\Z^T$ and every $i\in\{0,1,\dots,m-1\}$,
\begin{equation}\label{eq:incR}
(C\cdot Y)(u_i)=W\,Y(\beta_{i-1})-b_i\,Y(\alpha_i),\qquad (C\cdot Y)(v_i)=b_i\,Y(\alpha_i)-W\,Y(\beta_i).
\end{equation}

For $z\in\N$ define $Y_z\in\N^T$ by
\[
Y_z(\beta_i)=z,\qquad Y_z(\alpha_i)=\Big\lfloor\frac{\bar r_i+Wz}{b_i}\Big\rfloor\qquad(0\leq i\leq m-1),
\]
and let $D_z\coloneqq M_0+C\cdot Y_z$. Since $Y_z(\beta_{i-1})=Y_z(\beta_i)=z$, \eqref{eq:incR} gives, for
$0\leq i\leq m-1$,
\begin{equation}\label{eq:Dz}
\begin{aligned}
D_z(u_i)&=\bar r_i+Wz-b_i\big\lfloor(\bar r_i+Wz)/b_i\big\rfloor=(\bar r_i+Wz)\bmod b_i,\\
D_z(v_i)&=W-\bar r_i-Wz+b_i\big\lfloor(\bar r_i+Wz)/b_i\big\rfloor=W-D_z(u_i).
\end{aligned}
\end{equation}
So $D_z$ is the marking described in the paragraph above: the state equation gives it when every $\beta_i$ has
occurred $z$ times and every $\alpha_i$ as often as possible.

\begin{lemma}[key lemma]\label{lem:key}
Let $R=R(I,W)$ be as in Construction~\ref{con:ring}, and let $z\in\N$.
\begin{enumerate}[label=(\alph*),itemsep=1pt]
\item $D_z$ is a marking, and $D_0=M_0$. No $\alpha_i$ is enabled at $D_z$, and $\beta_i$ is enabled at $D_z$ iff
$Wz\equiv r_i\pmod{b_i}$. In particular, $D_z$ is dead iff $Wz$ is not covered by~$I$.
\item If $Wz$ is covered by $I$, then $D_z\fire{\sigma}D_{z+1}$ for some firing sequence $\sigma$ in which every
$\beta_j$ occurs exactly once.
\item $R$ is live if and only if $I$ covers $\Z$.
\item If $R$ is not live, then there is a least $z\in\N$ such that $Wz$ is not covered by $I$; call it $z^\ast$.
Every maximal firing sequence from $M_0$ is finite, has Parikh vector $Y_{z^\ast}$, and ends in the dead marking
$D_{z^\ast}$.
\end{enumerate}
\end{lemma}

\noindent\emph{Idea.} This is the counter of the paragraph ``How the ring works''. Part~(a) describes the markings
$D_z$, at which the counter shows $z$. Part~(b) lets the counter advance by one \emph{round}, in which the $W$
tokens moved by one $\beta_i$ are passed on around the ring, pair by pair. Part~(c) follows directly, and part~(d)
adds, with Corollary~\ref{cor:unique}, that every other order of firing ends in the same dead marking.

\begin{proof}
By~\eqref{eq:Dz}, for every $j\in\{0,1,\dots,m-1\}$ the place $u_j$ holds $D_z(u_j)=(\bar r_j+Wz)\bmod b_j$ tokens
at $D_z$, and $v_j$ holds $D_z(v_j)=W-D_z(u_j)$ tokens. Since $\bar r_j\equiv-r_j\pmod{b_j}$,
\begin{equation}\label{eq:zero}
D_z(u_j)=0\quad\iff\quad b_j\mid\bar r_j+Wz\quad\iff\quad Wz\equiv r_j\pmod{b_j}.
\end{equation}

\emph{(a)} Fix $j\in\{0,1,\dots,m-1\}$. We have $0\leq D_z(u_j)\leq b_j-1$, and hence
$D_z(v_j)=W-D_z(u_j)\geq W-b_j+1>0$, because $W>b_j$. So $D_z$ is a marking. For $z=0$, we have $Y_0(\beta_j)=0$
and $Y_0(\alpha_j)=\lfloor\bar r_j/b_j\rfloor=0$, because $0\leq\bar r_j<b_j$; so $Y_0=\zero$ and $D_0=M_0$. The
transition $\alpha_j$ is not enabled at $D_z$, because $u_j$ holds fewer than $b_j$ tokens. The transition $\beta_j$
is enabled at $D_z$ iff $v_j$ holds at least $W$ tokens, i.e.\ iff $W-D_z(u_j)\geq W$, i.e.\ iff $D_z(u_j)=0$; by
\eqref{eq:zero}, this means $Wz\equiv r_j\pmod{b_j}$. Hence some transition is enabled at $D_z$ iff $Wz$ lies in
one of the classes of $I$; that is, $D_z$ is dead iff $Wz$ is not covered.

\emph{(b)} Since $Wz$ is covered, there is an $i$ with $Wz\equiv r_i\pmod{b_i}$, and $\beta_i$ is enabled at $D_z$
by~(a). Starting at $D_z$, we fire the following \emph{round}~$\sigma$: first $\beta_i$; then, for the pairs
$j=i+1,i+2,\dots,i+m-1$ (indices modulo $m$) in this order, $\alpha_j$ as often as possible and then $\beta_j$ once;
finally $\alpha_i$ as often as possible. (For $m=1$, the round consists of $\beta_0$, followed by $\alpha_0$ as
often as possible.) Every $\beta_j$ occurs exactly once in $\sigma$.

Consider any pair $j$; its places are touched only by $\beta_{j-1}$, $\alpha_j$ and $\beta_j$. During the round,
$u_j$ receives $W$ tokens once (from $\beta_{j-1}$), $v_j$ gives away $W$ tokens once (by $\beta_j$), and $\alpha_j$
fires as often as possible after the $W$ tokens have arrived on $u_j$. Before they arrive, $u_j$ still holds its
$D_z(u_j)$ tokens, because only $\beta_{j-1}$ and $\alpha_j$ change the number of tokens on $u_j$, and $\alpha_j$
fires only after $\beta_{j-1}$. So the $\alpha_j$-firings leave on $u_j$ the
remainder of $D_z(u_j)+W$ modulo $b_j$, which is $(\bar r_j+W(z+1))\bmod b_j=D_{z+1}(u_j)$, because
$D_z(u_j)\equiv\bar r_j+Wz\pmod{b_j}$. The pair holds $W$ tokens at $D_z$ and has received and given away $W$
tokens each; so at the end, $v_j$ holds $W-D_{z+1}(u_j)=D_{z+1}(v_j)$ tokens. Hence $\sigma$ leads to $D_{z+1}$,
provided that all its firings are possible.

The $\alpha$-firings are possible by construction, and $\beta_i$ can fire at $D_z$ by~(a). For $j\neq i$, the
transition $\beta_j$ fires right after the $\alpha_j$-firings. At that moment the pair holds $2W$ tokens (its $W$
tokens at $D_z$ and the $W$ tokens received), of which only $D_{z+1}(u_j)<b_j<W$ lie on $u_j$; so $v_j$ holds more
than $W$ tokens, and $\beta_j$ is enabled. Hence $D_z\fire{\sigma}D_{z+1}$.

\emph{(c)} Suppose first that $I$ covers $\Z$. Then every $Wz$ is covered, and~(b) gives rounds
$D_0\fire{\sigma_0}D_1\fire{\sigma_1}D_2\fire{\sigma_2}\cdots$. Since $D_0=M_0$ by~(a), the infinite word
$\sigma_0\sigma_1\sigma_2\cdots$ is an infinite firing sequence from $M_0$, so $R$ is live by Lemma~\ref{lem:char},
(4)$\Rightarrow$(1).

Conversely, suppose that $I$ does not cover $\Z$, and let $x$ be an uncovered integer. Let
$B\coloneqq\lcm(b_0,b_1,\dots,b_{m-1})$. Every prime factor of $B$ divides some $b_i$, and hence does not divide $W$,
because $\gcd(W,b_i)=1$; so $\gcd(W,B)=1$. Therefore $W$ is invertible modulo $B$: there is an integer $W'$ with
$WW'\equiv1\pmod B$. Put $z\coloneqq(W'x)\bmod B$. Then $z\in\N$ (indeed $0\leq z<B$), and
\[
Wz\equiv WW'x\equiv x\pmod B.
\]
By the observation preceding~\eqref{eq:period}, $Wz$ is covered iff $x$ is; so $Wz$ is not covered. Hence the set
of all $z\in\N$ for which $Wz$ is not covered is non-empty; let $z^\ast$ be its least element. For every $z<z^\ast$
the number $Wz$ is covered, so~(b) gives rounds
$D_0\fire{\sigma_0}D_1\fire{\sigma_1}\cdots\fire{\sigma_{z^\ast-1}}D_{z^\ast}$, and $D_{z^\ast}$ is dead by~(a). So a
dead marking is reachable from $M_0$, and $R$ is not live.

This is where the assumption $\gcd(W,b_i)=1$ is needed: it guarantees that the multiples $Wz$, $z\in\N$, meet
every residue class modulo $B$. Without it, (c) can fail. Take $I=((0,2))$ and $W=4$, and build the ring as in
Construction~\ref{con:ring}, ignoring the condition $\gcd(W,b_0)=1$. Parts~(a) and~(b) do not use this condition.
Every $4z$ is even and hence covered, so by~(b) the counter advances forever, and this marked ring is live by
Lemma~\ref{lem:char}, although the odd integers are not covered.

\emph{(d)} Suppose that $R$ is not live. By~(c), $I$ does not cover $\Z$, and the proof of~(c) provides $z^\ast$
and the firing sequence $\rho\coloneqq\sigma_0\sigma_1\cdots\sigma_{z^\ast-1}$ from $M_0$ to the dead marking
$D_{z^\ast}$; so $\rho$ is a maximal firing sequence. Each $\beta_j$ occurs in $\rho$ exactly $z^\ast$ times, once in
each round. By the state equation and~\eqref{eq:incR}, for every $j\in\{0,1,\dots,m-1\}$,
\[
D_{z^\ast}(u_j)=M_0(u_j)+W\,\Par(\rho)(\beta_{j-1})-b_j\,\Par(\rho)(\alpha_j)=\bar r_j+Wz^\ast-b_j\,\Par(\rho)(\alpha_j).
\]
As $D_{z^\ast}(u_j)=(\bar r_j+Wz^\ast)\bmod b_j$, this gives
$\Par(\rho)(\alpha_j)=\lfloor(\bar r_j+Wz^\ast)/b_j\rfloor=Y_{z^\ast}(\alpha_j)$. Hence $\Par(\rho)=Y_{z^\ast}$. By
Corollary~\ref{cor:unique}, every maximal firing sequence from $M_0$ is finite, has the same Parikh vector
$\Par(\rho)=Y_{z^\ast}$, and ends in the same dead marking $D_{z^\ast}$.
\end{proof}

Part~(d) says that, whatever the order in which transitions occur, a non-live ring $R(I,W)$ always stops at the
same dead marking after the same numbers of occurrences: each $\beta_i$ occurs $z^\ast$ times and each $\alpha_i$
occurs $\lfloor(\bar r_i+Wz^\ast)/b_i\rfloor$ times. So the rounds of~(b) are only one way of running the ring;
other orders of firing, in which some $\beta_i$ may occur twice before another $\beta_j$ occurs at all, end in the
same dead marking (see Example~\ref{ex:twoinstances}).

\begin{figure}[t]
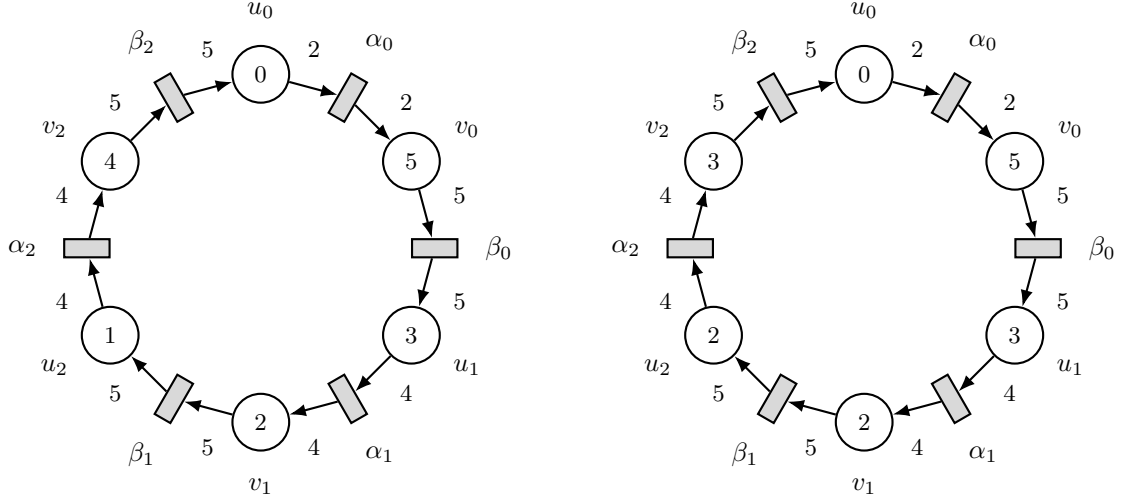

\centering
\begin{minipage}[t]{0.495\textwidth}\centering
\ringRIW{R=2.3,RL=3.15,RW=2.72,psize=7.5mm,theight=6mm,
  u0={$0$},v0={$5$},u1={$3$},v1={$2$},u2={$1$},v2={$4$},b0={$2$},b1={$4$},b2={$4$},W={$5$}}\\[1ex]
\small(a) $R(I_+,5)$, $I_+=((0,2),(1,4),(3,4))$: live
\end{minipage}\hfill
\begin{minipage}[t]{0.495\textwidth}\centering
\ringRIW{R=2.3,RL=3.15,RW=2.72,psize=7.5mm,theight=6mm,
  u0={$0$},v0={$5$},u1={$3$},v1={$2$},u2={$2$},v2={$3$},b0={$2$},b1={$4$},b2={$4$},W={$5$}}\\[1ex]
\small(b) $R(I_-,5)$, $I_-=((0,2),(1,4),(2,4))$: not live
\end{minipage}
\caption{The two rings of Example~\ref{ex:twoinstances}, drawn as in Figure~\ref{fig:ring}, with
$(b_0,b_1,b_2)=(2,4,4)$ and $W=5$; numbers inside places give the initial markings. The rings differ only in the
initial marking of $u_2$ and $v_2$. Ring~(a) is live. Ring~(b) is not live: every maximal firing sequence ends in
the dead marking $D_3$, which has $1,4,2,3,1,4$ tokens on $u_0,v_0,u_1,v_1,u_2,v_2$.}
\label{fig:examples}
\end{figure}

\begin{example}[a live and a non-live instance]\label{ex:twoinstances}
Let $W=5$ and consider the covering instances
\[
I_+=((0,2),(1,4),(3,4))\qquad\text{and}\qquad I_-=((0,2),(1,4),(2,4)),
\]
which differ only in the third pair. As noted after Definition~\ref{def:covering}, $I_+$ covers $\Z$. The instance
$I_-$ does not: the class $2\bmod4$ contains only even integers, which are covered by $0\bmod2$ anyway, so the
integers $x$ with $x\bmod4=3$ remain uncovered. Since $W=5$ is a prime larger than all moduli,
Construction~\ref{con:ring} applies, and $B=\lcm(2,4,4)=4$. We list markings in the order $u_0,v_0,u_1,v_1,u_2,v_2$.
We have $\bar r_0=(-0)\bmod2=0$ and $\bar r_1=(-1)\bmod4=3$ in both cases, while $\bar r_2=(-3)\bmod4=1$ for $I_+$
and $\bar r_2=(-2)\bmod4=2$ for $I_-$. So the initial markings are
\[
M_0^+=(0,5,3,2,1,4)\qquad\text{and}\qquad M_0^-=(0,5,3,2,2,3).
\]
Figure~\ref{fig:examples} shows both rings; they have the same weights and differ only in the initial marking of
$u_2$ and $v_2$.

By Lemma~\ref{lem:key}, everything depends on which of the numbers $Wz=5z$, $z\in\N$, are covered. Since
$5\equiv1\pmod 4$, we have $5z\equiv z\pmod4$, and since $B=4$, it suffices to consider $z=0,1,2,3$:
\begin{center}\small
\begin{tabular}{@{}ccccc@{}}
\toprule
$z$ & $5z$ & $5z\bmod4$ & covered in $I_+$ by & covered in $I_-$ by\\
\midrule
$0$ & $0$  & $0$ & $0\bmod2$ & $0\bmod2$\\
$1$ & $5$  & $1$ & $1\bmod4$ & $1\bmod4$\\
$2$ & $10$ & $2$ & $0\bmod2$ & $0\bmod2$ and $2\bmod4$\\
$3$ & $15$ & $3$ & $3\bmod4$ & --- (not covered)\\
\bottomrule
\end{tabular}
\end{center}
In the terms of the paragraph ``How the ring works'', the counter of $R(I_+,5)$ can advance forever, whereas the
counter of $R(I_-,5)$ advances from $0$ to $1$, $2$ and $3$ and gets stuck at $3$, because $15$ is not covered
by~$I_-$.

\emph{The live ring $R(I_+,5)$.} Every $5z$ is covered, so $R(I_+,5)$ is live by Lemma~\ref{lem:key}(c): its
counter advances forever. Liveness can also be seen directly. The least common
multiple of all weights is $L=\lcm(2,5,4,5,4,5)=20$, so by Lemma~\ref{lem:tinv} the minimal T-invariant
is given by $x_T(\alpha_0)=10$, $x_T(\alpha_1)=x_T(\alpha_2)=5$ and $x_T(\beta_i)=4$. The sequence
\[
\beta_0\,\alpha_1^2\,\beta_1^2\,\alpha_2^2\,\beta_2\,\alpha_0^2\,\beta_2\,\alpha_0^3\,\beta_0^2\,\alpha_1^2\,\beta_1\,
\alpha_2^2\,\beta_2\,\alpha_0^2\,\beta_2\,\alpha_0^3\,\beta_0\,\alpha_1\,\beta_1\,\alpha_2,
\]
where $\alpha^2$ stands for $\alpha\alpha$ and so on, is a firing sequence from $M_0^+$ with Parikh vector $x_T$;
it begins with $M_0^+\fire{\beta_0}(0,0,8,2,1,4)\fire{\alpha_1}(0,0,4,6,1,4)\fire{\alpha_1}(0,0,0,10,1,4)$. By
the state equation, it leads back to $M_0^+$, so repeating it forever gives an infinite firing sequence, and
$R(I_+,5)$ is live by Lemma~\ref{lem:char}, (4)$\Rightarrow$(1). (The reachability graph of $R(I_+,5)$, computed by
machine, has $74$ markings, none of which is dead.)

\emph{The non-live ring $R(I_-,5)$.} The least $z\in\N$ for which $5z$ is not covered by $I_-$ is $z^\ast=3$. By
Lemma~\ref{lem:key}(d), every maximal firing sequence from $M_0^-$ is finite; in it, each $\beta_i$ occurs exactly
$3$ times, and $\alpha_0$, $\alpha_1$, $\alpha_2$ occur $\lfloor15/2\rfloor=7$, $\lfloor(3+15)/4\rfloor=4$ and
$\lfloor(2+15)/4\rfloor=4$ times; and it ends in the dead marking $D_3$ with $D_3(u_i)=(\bar r_i+15)\bmod b_i$ and
$D_3(v_i)=5-D_3(u_i)$, i.e.
\[
D_3=(1,4,2,3,1,4).
\]
The rounds of Lemma~\ref{lem:key}(b) reach $D_3$ as follows. At $D_0=M_0^-$, the number $5\cdot0=0$ lies in the
class $0\bmod2$ of the first pair, so the first round starts with $\beta_0$:
\begin{align*}
D_0=(0,5,3,2,2,3)&\fire{\beta_0}(0,0,8,2,2,3)\fire{\alpha_1^2}(0,0,0,10,2,3)\fire{\beta_1}(0,0,0,5,7,3)\\
&\fire{\alpha_2}(0,0,0,5,3,7)\fire{\beta_2}(5,0,0,5,3,2)\fire{\alpha_0^2}(1,4,0,5,3,2)=D_1.
\end{align*}
Next, $5\cdot1=5$ lies in the class $1\bmod4$ of the second pair, and $5\cdot2=10$ in the class $0\bmod2$ of the
first pair (and also in $2\bmod4$); so the next two rounds can be
\[
D_1\fire{\beta_1\,\alpha_2^2\,\beta_2\,\alpha_0^3\,\beta_0\,\alpha_1}D_2=(0,5,1,4,0,5)
\fire{\beta_0\,\alpha_1\,\beta_1\,\alpha_2\,\beta_2\,\alpha_0^2}D_3.
\]
At $D_3$, each place $u_i$ holds a non-zero remainder ($1$, $2$ and $1$ tokens, fewer than $b_i=2,4,4$), so each
$v_i$ holds fewer than $W=5$ tokens, and no transition is enabled, in accordance with Lemma~\ref{lem:key}(a). Other
orders of firing lead to the same Parikh vector and the same dead marking, by Lemma~\ref{lem:key}(d). For instance,
firing in every step the first enabled transition in the order $\alpha_0,\beta_0,\alpha_1,\beta_1,\alpha_2,\beta_2$
yields the maximal firing sequence
\[
\beta_0\,\alpha_1^2\,\beta_1^2\,\alpha_2^3\,\beta_2\,\alpha_0^2\,\beta_2\,\alpha_0^3\,\beta_0^2\,\alpha_1^2\,\beta_1\,
\alpha_2\,\beta_2\,\alpha_0^2
\]
of length $24$, which also ends in $D_3$; in it, $\beta_1$ occurs twice before $\beta_2$ occurs at all. (The
reachability graph of
$R(I_-,5)$ has $93$ markings, and $D_3$ is the only dead one.)

In the live ring $R(I_+,5)$, the marking $D_3=(1,4,2,3,0,5)$ is not dead: there, $D_3(u_2)=(1+15)\bmod4=0$, so
$v_2$ holds all $5$ tokens of its pair and $\beta_2$ is enabled, as $15$ is covered by the class $3\bmod4$ of $I_+$.
\end{example}

\begin{theorem}[main result]\label{thm:main}
\pb{} is \coNP-complete. It is \coNP-hard already for instances $(N,M_0)$ in which $N$ is token-conserving and
all arc weights and all token numbers $M_0(s)$ are bounded by $O(n^{3/2}\log^3n)$, where $n$ is the number of
places; in particular, \pb{} is \coNP-hard in the strong sense.
\end{theorem}
\begin{proof}
Membership is Proposition~\ref{prop:upper}. For hardness, we reduce 3-SAT, restricted to formulas that have at
least one clause and no repeated clauses and whose clauses consist of three literals over three distinct variables
(which is still \NP-complete, see the proof of Proposition~\ref{prop:cover}), to the complement of \pb. Given such a
formula $\varphi$ with $k$ variables,
compute $I(\varphi)$ as in Proposition~\ref{prop:cover}, let $W$ be the least prime larger than all moduli of
$I(\varphi)$, and output the marked ring $R(I(\varphi),W)$.

\emph{Correctness.} Since $W$ is a prime larger than every modulus $b_i$ of $I(\varphi)$, we have
$\gcd(W,b_i)=1$, so Construction~\ref{con:ring} applies. By Proposition~\ref{prop:cover} and
Lemma~\ref{lem:key}(c),
\[
\varphi\text{ is unsatisfiable}\iff I(\varphi)\text{ covers }\Z\iff R(I(\varphi),W)\text{ is live}.
\]

\emph{Size and running time.} All moduli of $I(\varphi)$ are at most $p_k^3$, so by the Bertrand--Chebyshev
theorem (for every integer $h\geq1$ there is a prime $p$ with $h<p\leq2h$) we have $W\leq2p_k^3$; and $W$ can be found by
trial division in time polynomial in $p_k$, hence in $k$. The ring has $n=2m$ places, where
$m=l+\sum_{j=1}^k(p_j-2)$ by Proposition~\ref{prop:cover}. Since the $j$-th odd prime is at least $2j+1$, we have
$m\geq\sum_{j=1}^k(p_j-2)\geq\sum_{j=1}^k(2j-1)=k^2$, so $n\geq2k^2$, i.e.\ $k\leq\sqrt{n/2}$. Every arc weight
is some $b_i$ or $W$, hence at most $W$, and every place initially carries at most $W$ tokens ($M_0(u_i)=\bar r_i<b_i$ and
$M_0(v_i)=W-\bar r_i\leq W$). Since $p_k=O(k\log k)$, all these numbers are bounded by
\[
W\leq2p_k^3=O(k^3\log^3k)=O(n^{3/2}\log^3n).
\]
So \pb{} is \coNP-hard already for instances whose numbers are bounded by a polynomial in the number of places,
which is \coNP-hardness in the strong sense.
\end{proof}

\begin{remark}\label{rem:shape}
The hard instances are rather special: they are token-conserving, every second weight equals the same number $W$,
and no place initially carries more than $W$ tokens. As $W$ is a prime larger than all $b_i$, the least common
multiple of all weights is $L=W\cdot\lcm(b_0,b_1,\dots,b_{m-1})$, so the minimal T-invariant
(Lemma~\ref{lem:tinv}) has the entries $x_T(\alpha_i)=L/b_i$ and
\[
x_T(\beta_i)=L/W=\lcm(b_0,b_1,\dots,b_{m-1})=p_1p_2\cdots p_k ;
\]
the last equality, which expresses $x_T(\beta_i)$ as the product of the first $k$ odd primes, holds because every
modulus of $I(\varphi)$ is one of the primes $p_j$ or a product of three of them, and each $p_j$ occurs as a
modulus. These entries are superpolynomial but subexponential in $n$, although all numbers occurring in the
instance are polynomial in~$n$:
\begin{itemize}[leftmargin=1.4em,itemsep=1pt]
\item \emph{Lower bound.} $p_1p_2\cdots p_k\geq3^k$. The formula $\varphi$ has no repeated clauses (proof of
Theorem~\ref{thm:main}), so $l\leq8\binom k3$, hence $m=l+\sum_j(p_j-2)=O(k^3)$ and
$n=O(k^3)$, i.e.\ $k=\Omega(n^{1/3})$. Hence $x_T(\beta_i)\geq3^k=2^{\Omega(n^{1/3})}$.
\item \emph{Upper bound.} $p_1p_2\cdots p_k\leq p_k^k=2^{O(k\log k)}$, and $k\leq\sqrt{n/2}$ (proof of
Theorem~\ref{thm:main}). Hence $x_T(\beta_i)=2^{O(\sqrt n\log n)}$.
\end{itemize}
So a liveness test that has to fire every transition $t$ at least $x_T(t)$ times, such as the counterpart of the
criterion for plain T-systems mentioned in the Introduction, needs superpolynomially many firings on the live
instances, even though all numbers in them are small.
\end{remark}

\paragraph{Weighted T-systems.}
A \emph{weighted T-system}~\cite{TCCS92} is a net $N=(S,T,F)$ in which every place has exactly one input transition
and exactly
one output transition; the arc weights are arbitrary. An \emph{elementary circuit} $\gamma$ of $N$ is a sequence
$s_0\,t_0\,s_1\,t_1\cdots s_{h-1}\,t_{h-1}$ of $h\geq1$ distinct places $s_j$ and $h$ distinct transitions $t_j$ of
$N$ such that, for every $j$, the transition $t_j$ is the output transition of $s_j$ and the input transition of
$s_{j+1}$ (indices modulo $h$). It induces the ring $N_\gamma$ with the places $s_0,s_1,\dots,s_{h-1}$, the
transitions $t_0,t_1,\dots,t_{h-1}$ and the weights $a_j=F(s_j,t_j)$ and $b_j=F(t_j,s_{j+1})$, in the notation of
Section~\ref{sec:prelim}; a marking $M$ of $N$ induces the marking $M|_\gamma$ of $N_\gamma$, the restriction of $M$
to the places $s_j$. Every ring net is a weighted T-system; up to rotation, its only elementary circuit is the
whole ring.

\begin{lemma}[circuits; {\cite[Prop.~6.3]{HBDL20a}}]\label{lem:circuits}
A marked weighted T-system $(N,M_0)$ is live iff for every elementary circuit $\gamma$ of $N$, the marked ring
$(N_\gamma,M_0|_\gamma)$ is live.
\end{lemma}

\noindent\emph{Idea.} The places of a circuit are changed only by the transitions of the circuit, so every behaviour
of $N$ restricts to a behaviour of each ring $N_\gamma$. Conversely, if a transition can no longer occur, then one of
its input places never gets enough tokens again, so the input transition of this place also stops occurring, and
so on backwards, until a circuit closes.

\begin{proof}
We use two facts. (i) Firing a transition $t$ removes tokens only from places whose (only) output transition is
$t$; hence an enabled transition stays enabled until it fires. (ii) The input and output transitions of the places
of an elementary circuit $\gamma$ belong to $\gamma$. So, for every firing sequence $\sigma$ from $M_0$, the
subsequence $\sigma|_\gamma$ of the occurrences of transitions of $\gamma$ is a firing sequence of $N_\gamma$ from
$M_0|_\gamma$, which leads to the restriction to $\gamma$ of the marking reached by $\sigma$.

Suppose that $(N,M_0)$ is live. Firing the transitions in turn, which is possible by liveness, gives an infinite
firing sequence from $M_0$ in which every transition occurs infinitely often. By~(ii), its restriction to an
elementary circuit $\gamma$ is an infinite firing sequence of $N_\gamma$ from $M_0|_\gamma$, so
$(N_\gamma,M_0|_\gamma)$ is live by Lemma~\ref{lem:char}, (4)$\Rightarrow$(1).

Conversely, suppose that $(N,M_0)$ is not live: some transition $t$ occurs in no firing sequence from some
reachable marking $M$. Let $\rho$ be a maximal firing sequence from $M$ that is \emph{fair}, i.e.\ every transition
that is enabled at some point of $\rho$ occurs at a later point of $\rho$. Such a $\rho$ exists by~(i): keep the
enabled transitions in a queue; repeatedly fire the first one, remove it from the queue, and append every enabled
transition that is not yet in the queue. Let $\tau$ be a transition that occurs only finitely often in $\rho$, for
instance $t$. After its last occurrence, $\tau$ is never enabled again, by fairness. Only $\tau$ removes tokens from
its input places, so their token numbers do not decrease after this point. Hence some input place $q$ of $\tau$
carries fewer than $F(q,\tau)$ tokens from some point of $\rho$ on (otherwise, from some point on, all input places
of $\tau$ would carry enough tokens, and $\tau$ would be enabled). Each occurrence of the input transition of $q$
adds tokens to $q$, so this transition, too, occurs only finitely often in $\rho$. Starting with $\tau_0\coloneqq t$ and repeating this
step, we obtain transitions $\tau_0,\tau_1,\tau_2,\dots$ that occur only finitely often in $\rho$, and places
$q_0,q_1,q_2,\dots$ such that $q_i$ has the output transition $\tau_i$ and the input transition $\tau_{i+1}$ and
carries fewer than $F(q_i,\tau_i)$ tokens from some point of $\rho$ on. Since $T$ is finite, there is a least $b$ such
that $\tau_b=\tau_a$ for some $a<b$. Then $\tau_a,\tau_{a+1},\dots,\tau_{b-1}$ are distinct, and so are
$q_a,q_{a+1},\dots,q_{b-1}$, whose output transitions differ; hence
$\gamma=q_{b-1}\,\tau_{b-1}\,q_{b-2}\,\tau_{b-2}\cdots q_a\,\tau_a$ is an elementary circuit of $N$. From some point of
$\rho$ on, every place of $\gamma$ carries fewer tokens than its output transition needs. By~(ii), the marking reached
at this point, which is reachable from $M_0$, restricts to a dead marking of $N_\gamma$ that is reachable from
$M_0|_\gamma$. So $(N_\gamma,M_0|_\gamma)$ is not live.
\end{proof}

\begin{corollary}[weighted T-systems]\label{cor:wts}
Liveness of weighted T-systems is \coNP-complete; it is \coNP-hard in the strong sense already for
token-conserving rings.
\end{corollary}
\begin{proof}
A marked weighted T-system is given by its places, transitions, arc weights and initial marking, with numbers in
binary. Hardness holds by Theorem~\ref{thm:main}, since ring nets are weighted T-systems. For membership, let
$(N,M_0)$ be a marked weighted T-system that is not live. By Lemma~\ref{lem:circuits}, some elementary circuit
$\gamma$ of $N$ induces a marked ring $(N_\gamma,M_0|_\gamma)$ that is not live. All numbers of this ring occur in
$(N,M_0)$, so, by the proof of Proposition~\ref{prop:upper}, there is a vector $Y$ whose entries have polynomially
many bits such that $M_0|_\gamma+C_\gamma\cdot Y$ is a dead marking of $N_\gamma$, where $C_\gamma$ is the incidence
matrix of $N_\gamma$. The pair $(\gamma,Y)$ is a certificate of polynomial size for non-liveness: one can check in
polynomial time that $\gamma$ is an elementary circuit and that $M_0|_\gamma+C_\gamma\cdot Y$ is a dead marking of
$N_\gamma$, and then $(N_\gamma,M_0|_\gamma)$ is not live by Lemma~\ref{lem:char}, (1)$\Rightarrow$(6), and
$(N,M_0)$ is not live by Lemma~\ref{lem:circuits}.
\end{proof}

\section{Chip-firing on simple digraphs}\label{sec:chip}

Chip-firing is a solitaire game on a graph, introduced for undirected graphs by Björner, Lovász and
Shor~\cite{BLS91} and studied on directed graphs by Björner and Lovász~\cite{BL92}. In this section we show that
its halting problem is \NP-complete on simple directed graphs, and \NP-hard already on strongly connected ones, by a
reduction from non-liveness of token-conserving rings. We first recall the definitions.

\paragraph{Chip-firing.}
Let $G=(V,E)$ be a finite directed graph (\emph{digraph}) with vertex set $V$ and edge set $E\subseteq V\times V$.
Such a digraph is \emph{simple}: since $E$ is a set, every edge occurs at most once; we also exclude loops $(v,v)$.
(Farrell and Levine~\cite{FL16} also consider \emph{multigraphs}, in which an edge may occur several times.) The
\emph{out-degree} $d(v)$ of a vertex $v$ is the number of edges $(v,w)\in E$; the vertices $w$ with $(v,w)\in E$ are
the \emph{out-neighbours} of $v$, and the vertices $y$ with $(y,v)\in E$ are its \emph{in-neighbours}. A \emph{chip
configuration} is a vector $\sigma\in\N^V$; we think of $\sigma(v)$ as the number of \emph{chips} on $v$. A vertex
$v$ is \emph{active} at $\sigma$ if $d(v)\geq1$ and $\sigma(v)\geq d(v)$; thus a vertex without outgoing edges
never fires (in a strongly connected digraph with at least two vertices, the setting of~\cite{FL16}, there are no
such vertices). An active vertex $v$ may \emph{fire}: it
sends one chip along each of its outgoing edges, i.e.\ it loses $d(v)$ chips, and each of its out-neighbours gains one
chip. A \emph{legal game} from $\sigma$ is a finite or infinite sequence of firings, each of which fires a vertex
that is active at the current configuration. A configuration is \emph{stable} if no vertex is active at it, and
$\sigma$ \emph{halts} if some legal game from $\sigma$ ends in a stable configuration. The
\textup{\textsc{Halting Problem for Chip-Firing}} asks, given $G$ and $\sigma$ (with numbers in binary), whether
$\sigma$ halts.

In Petri net terms, chip-firing on $G$ is the net whose places are the vertices of $G$ and which has a transition
$f_v$ for every vertex $v$ with $d(v)\geq1$; the transition $f_v$ takes $d(v)$ tokens from $v$ and puts one token on
each out-neighbour of $v$. Chip configurations are markings, legal games are firing sequences, stable configurations
are dead markings, and $\sigma$ halts iff a dead marking is reachable from $\sigma$.

\paragraph{What was known.}
Björner and Lovász~\cite{BL92} showed that if some legal game from $\sigma$ ends in a stable configuration, then no
legal game from $\sigma$ is infinite, and all legal games from $\sigma$ that cannot be continued end in the same
stable configuration, after the same number of firings of each vertex. (For rings, Corollary~\ref{cor:unique} is the
analogous statement.) So it does not matter which active vertex is fired: if $\sigma$ halts, every legal game from
$\sigma$ stops after finitely many firings. Farrell and Levine~\cite[Cor.~3.2]{FL16} proved that the halting problem
is \NP-complete for strongly connected directed multigraphs; their hard instances may have large edge multiplicities.
They asked ``whether the \textup{\textsc{Halting Problem for Chip-Firing}} remains \NP-hard when restricted to simple
directed graphs''~\cite[Sect.~3.1]{FL16}. The question is also posed as an open problem in the EGRES open problem
collection~\cite{EGRESp}, whose overview page on chip-firing~\cite{EGRES} conjectures that the problem is \NP-hard
and records that it is polynomial for simple Eulerian digraphs. See also~\cite{BLS91,Tot22}.

\paragraph{The reduction.}
The idea is to let each place $s_i$ of a token-conserving ring become a vertex $v_i$, and to replace the
transition $t_i$, which moves $A_i$ tokens from $s_i$ to $s_{i+1}$, by $A_i$ parallel paths of length two from
$v_i$ to $v_{i+1}$.

\begin{construction}[the digraph $G_N$]\label{con:chip}
Let $N$ be a token-conserving ring with weights $A_i\coloneqq a_i=b_i$ ($0\leq i\leq n-1$), and let $M_0$ be a
marking of $N$. The digraph $G_N$ has the vertices
\[
v_0,v_1,\dots,v_{n-1}\qquad\text{and}\qquad w_{i,k}\quad(0\leq i\leq n-1,\ 1\leq k\leq A_i),
\]
and the edges $v_i\to w_{i,k}$ and $w_{i,k}\to v_{i+1}$ for all $i$ and $k$, where $v_n\coloneqq v_0$. The chip
configuration $\sigma_{M_0}$ puts $M_0(s_i)$ chips on $v_i$, for every $i$, and no chips on the vertices $w_{i,k}$.
\end{construction}

\begin{figure}[t]
\centering
\begin{tikzpicture}[
  hub/.style={circle,draw,thick,minimum size=9mm,inner sep=0pt},
  sub/.style={circle,draw,thick,minimum size=8.5mm,inner sep=0pt,font=\scriptsize},
  e/.style={-{Latex[length=2mm]},thick}]
  \node[hub,label=left:{$v_0$}] (v0) at (0,0) {$3$};
  \node[hub,label=right:{$v_1$}] (v1) at (6,0) {$0$};
  \node[sub] (w01) at (3,2.0) {$w_{0,1}$};
  \node[sub] (w02) at (3,1.0) {$w_{0,2}$};
  \node[sub] (w11) at (3,-1.0) {$w_{1,1}$};
  \node[sub] (w12) at (3,-2.0) {$w_{1,2}$};
  \node[sub] (w13) at (3,-3.0) {$w_{1,3}$};
  \foreach \w in {w01,w02} {\draw[e] (v0) -- (\w); \draw[e] (\w) -- (v1);}
  \foreach \w in {w11,w12,w13} {\draw[e] (v1) -- (\w); \draw[e] (\w) -- (v0);}
\end{tikzpicture}
\caption{The digraph $G_N$ of Construction~\ref{con:chip} for the token-conserving ring with $n=2$ and weights
$A_0=2$, $A_1=3$ (Example~\ref{ex:chip}). The numbers inside $v_0$ and $v_1$ are their chips in $\sigma_{M_0}$ for
$M_0=(3,0)$; the vertices $w_{i,k}$ carry no chips. Firing $v_0$ and then $w_{0,1}$ and $w_{0,2}$ has the same effect
as one occurrence of $t_0$.}
\label{fig:chip}
\end{figure}
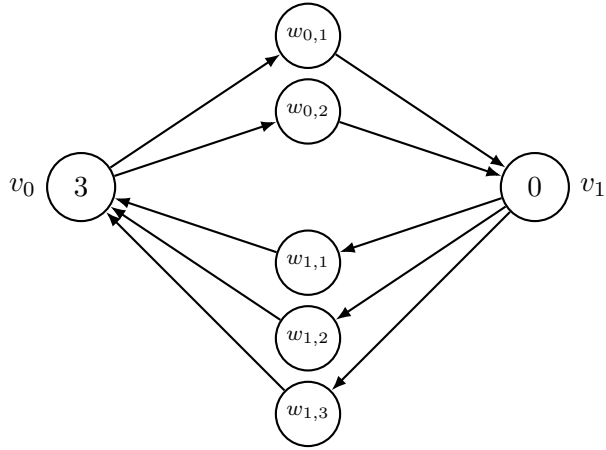

The vertex $v_i$ plays the role of the place $s_i$, and the $A_i$ paths $v_i\to w_{i,k}\to v_{i+1}$ play the role
of the transition $t_i$: firing $v_i$ sends $A_i$ chips into these paths, one into each, and firing the vertices
$w_{i,k}$ delivers them to $v_{i+1}$. The digraph $G_N$ is simple: it has no loops, and no edge occurs twice, because
the vertices $w_{i,k}$ are distinct. It is strongly connected, since the paths lead all around the ring. It has
$n+\sum_iA_i$ vertices and $2\sum_iA_i$ edges, and the out-degrees are $d(v_i)=A_i$ and $d(w_{i,k})=1$.
Figure~\ref{fig:chip} shows an example.

\begin{proposition}\label{prop:chip}
Let $N$ be a token-conserving ring and $M_0$ a marking of $N$. Then $\sigma_{M_0}$ halts on $G_N$ iff $M_0$ is not
live.
\end{proposition}
\begin{proof}
Let $A_i$ be the weights of $N$. By~\eqref{eq:incidence}, $(C\cdot Y)(s_i)=A_{i-1}Y_{i-1}-A_iY_i$ for every
$Y\in\Z^n$.

\emph{If $M_0$ is not live, then $\sigma_{M_0}$ halts.} By Lemma~\ref{lem:char}, some firing sequence $\rho$ leads from
$M_0$ to a dead marking $D$. We translate $\rho$ into a legal game: every occurrence of a transition $t_i$ in $\rho$
is replaced by a \emph{block} of firings, namely a firing of $v_i$, followed by one firing of each of the vertices
$w_{i,1},w_{i,2},\dots,w_{i,A_i}$. We claim that before and after each block, every vertex $v_j$ carries as many chips as
$s_j$ carries tokens at the current marking of $N$, and the vertices $w_{j,k}$ carry no chips. This holds at the
start, by the definition of $\sigma_{M_0}$. Suppose that it holds before the block for an occurrence of $t_i$ at a
marking $M$. Since $t_i$ is enabled at $M$, the vertex $v_i$ carries $M(s_i)\geq A_i=d(v_i)$ chips, so it is
active. Firing it leaves $M(s_i)-A_i$ chips on $v_i$ and puts one chip on each $w_{i,k}$. Then each $w_{i,k}$ is
active, as $d(w_{i,k})=1$, and firing it moves its chip to $v_{i+1}$. After the block, $v_i$ has lost $A_i$ chips,
$v_{i+1}$ has gained $A_i$ chips, and the vertices $w_{i,k}$ are empty again; this is exactly the effect of $t_i$ on
the marking. (For $n=1$, $v_{i+1}=v_i$, and $v_0$ loses and regains $A_0$ chips, just as $s_0$ loses and regains
$A_0$ tokens.) So at the end of the game the configuration is $D$ on the vertices $v_j$ and $0$ on the vertices
$w_{j,k}$. It is stable: $v_j$ carries $D(s_j)<A_j=d(v_j)$ chips, because $D$ is dead, and $w_{j,k}$ carries
$0<1=d(w_{j,k})$ chips. Hence $\sigma_{M_0}$ halts.

\emph{If $\sigma_{M_0}$ halts, then $M_0$ is not live.} Let a legal game from $\sigma_{M_0}$ end in a stable configuration
$\tau$, and for every vertex $x$ let $u(x)$ be the number of firings of $x$ in this game. Each firing of $x$ removes
$d(x)$ chips from $x$, and each firing of an in-neighbour of $x$ adds one chip to $x$; so, by counting,
\[
\tau(x)=\sigma_{M_0}(x)-d(x)\,u(x)+\sum_{y\colon(y,x)\in E}u(y)\qquad\text{for every vertex }x.
\]
(This is the state equation~\eqref{eq:stateeq} of the Petri net described above.) The only in-neighbour of
$w_{i,k}$ is $v_i$, so $\tau(w_{i,k})=0-u(w_{i,k})+u(v_i)$. Since $\tau$ is stable, $\tau(w_{i,k})<d(w_{i,k})=1$,
i.e.\ $\tau(w_{i,k})=0$, and hence $u(w_{i,k})=u(v_i)$: in total, each $w_{i,k}$ has passed on every chip it has
received. The in-neighbours of $v_i$ are the vertices $w_{i-1,k}$, $1\leq k\leq A_{i-1}$, each of which fires
$u(v_{i-1})$ times. So, with $Y_i\coloneqq u(v_i)$ for $0\leq i\leq n-1$,
\[
\tau(v_i)=M_0(s_i)-A_iY_i+A_{i-1}Y_{i-1}=(M_0+C\cdot Y)(s_i).
\]
Since $\tau$ is a stable configuration, $0\leq\tau(v_i)<d(v_i)=A_i$ for all $i$. So $Y\in\N^n$, and $M_0+C\cdot Y$ is
a dead marking of $N$. By Lemma~\ref{lem:char}, (1)$\Leftrightarrow$(6), $M_0$ is not live.
\end{proof}

The second part of the proof does not translate the game back into a firing sequence of $N$ step by step. This would
not be straightforward: in a legal game, a vertex $w_{i,k}$ may keep its chip for a long time while other vertices
fire. Counting how often each vertex fires avoids this difficulty, just as the state equation records only how often
each transition occurs.

\begin{example}\label{ex:chip}
Let $N$ be the token-conserving ring with $n=2$ and weights $A_0=2$, $A_1=3$; its digraph $G_N$ is shown in
Figure~\ref{fig:chip}. For $M_0=(3,0)$, the only active vertex is $v_0$. Firing it leaves $1$ chip on $v_0$ and puts
$1$ chip on each of $w_{0,1},w_{0,2}$; these two vertices fire, and $v_1$ receives $2$ chips. Now $v_0$ has $1<2$
chips, $v_1$ has $2<3$ chips, and all vertices $w_{i,k}$ are empty, so the configuration is stable and $\sigma_{M_0}$
halts. This mirrors $(3,0)\fire{t_0}(1,2)$, where $(1,2)$ is a dead marking of $N$. The marking $(2,2)$ is
live, since
\[
(2,2)\fire{t_0}(0,4)\fire{t_1}(3,1)\fire{t_0}(1,3)\fire{t_1}(4,0)\fire{t_0}(2,2)
\]
returns to $(2,2)$; correspondingly, for $M_0=(2,2)$ the configuration $\sigma_{M_0}$ does not halt, and every legal game from $\sigma_{M_0}$ can
be continued forever.
\end{example}

\paragraph{Membership in \NP.}
For a digraph $G=(V,E)$ and a vector $u\in\N^V$, let $\Delta u\in\Z^V$ be given by
$(\Delta u)(x)\coloneqq d(x)\,u(x)-\sum_{y\colon(y,x)\in E}u(y)$. By the counting argument in the proof of
Proposition~\ref{prop:chip}, a legal game from $\sigma$ in which every vertex $x$ fires $u(x)$ times leads to the
configuration $\sigma-\Delta u$. The following lemma is the chip-firing counterpart of Lemma~\ref{lem:bound}; it is
known as the least action principle (cf.~\cite[Lemma~2.3]{FLP10} for sandpiles on $\Z^d$, and~\cite{BoL16} for
abelian networks, which include chip-firing on digraphs).

\begin{lemma}[least action principle]\label{lem:leastaction}
Let $\sigma$ be a chip configuration on a digraph $G=(V,E)$, and let $u\in\N^V$ be such that
$(\sigma-\Delta u)(x)<d(x)$ for every vertex $x$ with $d(x)\geq1$; negative entries of $\sigma-\Delta u$ are allowed.
Then in every legal game from $\sigma$, every vertex $x$ fires at most $u(x)$ times. Consequently, $\sigma$ halts iff
such a vector $u$ exists.
\end{lemma}
\begin{proof}
Suppose that in some legal game from $\sigma$ some vertex fires more often than this, and consider the first firing
in it at which a vertex $x$ fires for the $(u(x)+1)$-st time. Before this firing, $x$ has fired $u(x)$ times
and every in-neighbour $y$ of $x$ at most $u(y)$ times, so $x$ carries at most
$\sigma(x)-d(x)\,u(x)+\sum_{y\colon(y,x)\in E}u(y)=(\sigma-\Delta u)(x)<d(x)$ chips and is not active, a
contradiction. So if such a vector $u$ exists, every legal game from $\sigma$ is finite; hence firing active vertices
as long as possible leads to a stable configuration, and $\sigma$ halts. Conversely, if a legal game from $\sigma$
ends in a stable configuration $\tau$, let $u(x)$ be the number of firings of $x$ in it; then $\tau=\sigma-\Delta u$,
and $\tau(x)<d(x)$ for every vertex $x$ with $d(x)\geq1$.
\end{proof}

\begin{corollary}\label{cor:chip}
The \textup{\textsc{Halting Problem for Chip-Firing}} is \NP-complete on simple digraphs. It is \NP-hard in the
strong sense already for strongly connected digraphs, namely for the digraphs $G_N$ of Construction~\ref{con:chip},
in which all but $n$ vertices have out-degree~$1$, and for configurations whose numbers of chips are bounded by a
polynomial in the number of vertices.
\end{corollary}
\begin{proof}
\emph{Membership in \NP.} By Lemma~\ref{lem:leastaction}, $\sigma$ halts on $G=(V,E)$ iff there is $u\in\N^V$ with
$(\sigma-\Delta u)(x)\leq d(x)-1$ for every vertex $x$ with $d(x)\geq1$. With a slack variable for each of these
inequalities, this becomes a system of linear equations with non-negative integer unknowns, as in the proof of
Proposition~\ref{prop:upper}; it has at most $|V|$ equations and $2|V|$ unknowns, and all its coefficients and
right-hand sides have absolute value at most $|V|+\max_x\sigma(x)$. By the theorem of
Papadimitriou~\cite[Theorem, p.~767]{Pap81}, if the system has a solution, it has one whose entries have polynomially
many bits. Such a $u$ can be guessed and checked in polynomial time. (For strongly connected multigraphs, membership
in \NP{} is~\cite[Cor.~3.2]{FL16}.)

\emph{Hardness.} We reduce 3-SAT, as in the proof of Theorem~\ref{thm:main}. Given a formula $\varphi$, let
$(N,M_0)$ be the marked ring $R(I(\varphi),W)$ constructed there; then $\varphi$ is satisfiable iff $M_0$ is not
live. All weights $A_i$ of $N$ and all token numbers $M_0(s)$ are bounded by a polynomial in the number $n$ of places,
and $n$ is polynomial in the size of $\varphi$. Hence $G_N$, which has $n+\sum_iA_i$ vertices, and $\sigma_{M_0}$ can
be computed in polynomial time, and the numbers of chips in $\sigma_{M_0}$ are polynomially bounded. By
Proposition~\ref{prop:chip}, $\sigma_{M_0}$ halts on $G_N$ iff $M_0$ is not live, i.e.\ iff $\varphi$ is satisfiable.
\end{proof}

This answers the question of Farrell and Levine and the question in~\cite{EGRESp}, and confirms the conjecture
recorded in~\cite{EGRES}. Note the role
of strong hardness in Theorem~\ref{thm:main}: replacing a weight $A_i$ by $A_i$ parallel paths takes polynomial time
only because the weights are polynomially bounded.

\section{Concluding remarks}\label{sec:conclusion}

Liveness of weighted T-systems is \coNP-complete, and it is \coNP-hard in the strong sense already for single
circuits, even when they are token-conserving; so, unless $\Ptime=\NP$, it can be
decided neither by a polynomial nor by a pseudo-polynomial algorithm. As a consequence, the halting problem for
chip-firing is \NP-complete on simple digraphs. A later version of this paper is planned to contain further results
on reversibility and reachability of ring nets and on tractable special cases, as well as details of the
computational cross-checks mentioned in the Introduction.

\end{document}